\documentclass[12pt]{article}
\usepackage{amsmath,
			amssymb,
			amsthm,
			amscd,
			amsfonts,
			amstext,
			indentfirst,
			color,
			slashed,
                   cite}

\usepackage{eucal,dsfont,enumitem}
\usepackage[all]{xy}
\usepackage[active]{srcltx}
\setlist{nolistsep}

\usepackage[debug,pageanchor=false]{hyperref}
\hypersetup{colorlinks=true,
			linktocpage,
			breaklinks,
		 	urlcolor=blue,
		 	linkcolor=red,
		 	citecolor=blue}

\usepackage[top=0.7in,
			bottom=1in,
			left=1in,
			right=1in,
			includehead]
			{geometry}
			
\usepackage{tikz}
\usetikzlibrary{positioning,arrows}
\tikzset{
			>=stealth',
			format/.style={
					  rectangle,
					  rounded corners,
                      minimum size=2cm,
                      draw=black,very thick,
                      font=\fontsize{12}{14}\selectfont},
			arrow/.style={
					  ->,
					  thick,
					  shorten <=2pt,
					  shorten >=2pt,}
		}

\numberwithin{equation}{section}

\usepackage{setspace}
\def\be{\begin{equation}}
\def\ee{\end{equation}}
\def\ben{\begin{equation*}}
\def\een{\end{equation*}}
\def\ba{\begin{array}}
\def\ea{\end{array}}
\def\bn{\begin{aligned}}
\def\en{\end{aligned}}
\def\bg{\begin{gathered}}
\def\eg{\end{gathered}}
\def\bsub{\begin{subequations}}
\def\esub{\end{subequations}}

\def\bea{\begin{eqnarray}}
\def\eea{\end{eqnarray}}

\def\p{{\partial}}

\def\5{AdS_5 \times S_5}

\def\d{{\delta}}

\def\e{{\epsilon}}

\def\h{{\eta}}

\def\Th{{\Theta}}

\def\LL{{\Lambda}}
\def\m{{\mu}}
\def\n{{\nu}}

\def\S{{\Sigma}}

\def\O{{\Omega}}

\newcommand{\adss}{AdS$_5\times$S$^5$ }

\def\Dslash{\,\,{\raise.15ex\hbox{/}\mkern-12mu D}}
\def\Dbarslash{\,\,{\raise.15ex\hbox{/}\mkern-12mu {\bar D}}}
\def\delslash{\,\,{\raise.15ex\hbox{/}\mkern-9mu \partial}}
\def\delbarslash{\,\,{\raise.15ex\hbox{/}\mkern-9mu {\bar\partial}}}
\def\pslash{\,\,{\raise.15ex\hbox{/}\mkern-9mu p}}
\def\calDslash{\,\,{\raise.15ex\hbox{/}\mkern-12mu {\cal D}}}

\newcommand\diff{\mbox{d}}

\newcommand{\vol}{\mbox{vol}}
\newcommand{\bse}{\begin{subequations}}
\newcommand{\ese}{\end{subequations}}

\newcommand{\dd}{\diff}

\newcommand{\bd}{{\rm d}}
\newcommand{\no}{\nonumber}
\newcommand{\mL}{\mathcal L}
\newcommand{\nn}{\nonumber \\}

\begin{document}

\begin{titlepage}

\begin{flushright}\vspace*{-3cm}
{\small
%{\tt arXiv:yymm.nnnn} \\
IPM/P-2017/016,
{APCTP Pre2017-004},
{KUNS-2676}\\
%\today
}
\end{flushright}
\vspace{1cm}

\begin{center}
{\Large{\bf{  Conformal Twists, Yang-Baxter $\sigma$-models \vspace*{3.0mm} \\
\& Holographic Noncommutativity }}}
\vspace{15mm}

\large{{Thiago Araujo\footnote{e-mail:~\href{mailto:thiago.araujo@apctp.org}{thiago.araujo@apctp.org}}$^{; a}$, Ilya Bakhmatov\footnote{email:~\href{ilya.bakhmatov@apctp.org}{ilya.bakhmatov@apctp.org}}$^{; a,b}$, Eoin \'O Colg\'ain\footnote{e-mail:~\href{ocolgain.eoin@apctp.org}{ocolgain.eoin@apctp.org}}$^{; a}$},  Jun-ichi  Sakamoto\footnote{e-mail:~\href{sakajun@gauge.scphys.kyoto-u.ac.jp}{sakajun@gauge.scphys.kyoto-u.ac.jp}}$^{; c}$, Mohammad M.~Sheikh-Jabbari\footnote{e-mail:~\href{jabbari@theory.ipm.ac.ir}{jabbari@theory.ipm.ac.ir}}$^{; d}$,  {Kentaroh Yoshida}\footnote{e-mail:~\href{kyoshida@gauge.scphys.kyoto-u.ac.jp}{kyoshida@gauge.scphys.kyoto-u.ac.jp}}$^{; c}$}
\\

\vspace*{5mm}

\normalsize
\bigskip

 $^a$\textit{Asia Pacific Center for Theoretical Physics, Postech, Pohang 37673, Korea}\\
\vspace{1.7mm}
 $^b$\textit{Institute of Physics, Kazan Federal
University, Kremlevskaya 16a, 420111, Kazan, Russia}\\
\vspace{1.7mm}
$^c$\textit{Department of Physics, Kyoto University,  Kitashirakawa, Kyoto 606-8502, Japan}\\
\vspace{1.7mm}
$^d$\textit{School of Physics, Institute for Research in Fundamental Sciences (IPM), Tehran, Iran}
\\

\vspace{15mm}

\begin{abstract}
Expanding upon earlier results [arXiv:1702.02861], we present a compendium of  $\sigma$-models associated with integrable deformations of AdS$_5$ generated by solutions to homogenous classical Yang-Baxter equation. Each example we study from four viewpoints: conformal (Drinfeld) twists, closed string gravity backgrounds, open string parameters and proposed dual noncommutative (NC) gauge theory. Irrespective of whether the deformed background is a solution to supergravity or generalized supergravity, we show that  the open string metric associated with each gravity background is undeformed AdS$_5$ with constant open string coupling and the NC structure $\Theta$ is directly related to the conformal twist. One novel feature is that $\Theta$ exhibits ``holographic noncommutativity'': while it may exhibit non-trivial dependence on the holographic direction, its value everywhere in the bulk is uniquely determined by its value at the boundary, thus facilitating introduction of a dual NC gauge theory. We show that the divergence of the NC structure $\Theta$ is directly related to the unimodularity of the twist. We discuss the implementation of an outer automorphism of the conformal algebra as a coordinate transformation in the AdS bulk and discuss its implications for Yang-Baxter $\sigma$-models and self-T-duality based on fermionic T-duality. Finally, we comment on implications of our results for the integrability of associated open strings and planar integrability of dual NC gauge theories.

\end{abstract}

%\pacs{04.65.+e,04.70.-s,11.30.-j,12.10.-g}

\end{center}
%%%%%%%%%%%%%%%%%%%%%%%%%%%%%%%%%%%%%%%%%%%%%%%%%%%%%%%%%%%%%%%%%%%%%%%%%%%%%%%%%%%%%%%%

\end{titlepage}
%%%%%%%%%%%%%%%%%%%%%%%%%%%%%%%%%%%%%%%%%%%%%%%%%%%%%%%%%%%%%%%%%%%%%%%%%%%%%%%%%%%%%%%%%%
\setcounter{footnote}{0}
\setstretch{1.3}
%%%%%%%%%%%%%%%%%%%%%%%%%%%%%%%%%%%%%%%%%%%%%%%%%%%%%%%%%%%%%%%%%%%%%%%%%%%%%%%%
{\footnotesize
\tableofcontents
}
\section{Introduction}
The study of integrable models has precipitated many powerful results in physics, leading to deep insights, especially in non-perturbative regimes. String theory is no exception: in the wake of the discovery that the two-dimensional (2D) string worldsheet $\sigma$-model on \adss is classically integrable \cite{Bena:2003wd}, great interest has focused on integrable structures at the heart of the AdS/CFT correspondence \cite{Maldacena:1997re}. More concretely, we recognize that the string worldsheet theory is classically integrable in the sense that one can define a \textit{Lax pair} (see \cite{Arutyunov:2009ga,Beisert:2010jr,Torrielli:2016ufi} for reviews), whose flatness is equivalent to the equations of motion of the $\sigma$-model. In turn, the Lax pair allows one to define an infinite set of conserved charges. In the context of the prototypical AdS/CFT duality between string theory on \adss and $\mathcal{N}=4$ super Yang-Mills (sYM), integrability has provided the strongest evidence for the duality, allowing one in principle to compute the spectrum of the quantum theory in the planar limit exactly \cite{Bombardelli:2009ns, Gromov:2009bc, Arutyunov:2009ur, Gromov:2013pga}.

There is considerable interest in extending integrability techniques beyond the highly symmetric setting of \adss, or equivalently sYM on $\mathbb{R}^{1,3}$, to less contrived examples. We recall that one of the earliest integrability preserving deformations of \adss was discovered in the process of constructing AdS/CFT geometries  \cite{Hashimoto:1999ut, Maldacena:1999mh, Alishahiha:1999ci} dual to NC  spacetimes, which arise naturally in string theory \cite{Ardalan:1998ce, Seiberg:1999vs} (see \cite{Szabo:2001kg} for a review). In time we came to appreciate these deformations as a particular example of T-duality shift T-duality (TsT) transformations \cite{Lunin:2005jy}. For TsT deformations, the existence of a Lax pair guarantees that the deformation is classically integrable \cite{Frolov:2005dj} (see also \cite{Frolov:2005ty})

Recently, Yang-Baxter (YB) deformations of the $\sigma$-model \cite{Klimcik:2002zj, Klimcik:2008eq, Delduc:2013fga, Matsumoto:2015jja} have emerged as a systematic way to generate integrable deformations of the \adss superstring \cite{Delduc:2013qra, Kawaguchi:2014qwa}. As a result, one can embed TsT transformations into a larger class of YB deformations of the $\sigma$-model \cite{Matsumoto:2014nra, Matsumoto:2014gwa,Matsumoto:2015uja, Matsumoto:2014ubv,Kawaguchi:2014fca,vanTongeren:2015soa,vanTongeren:2015uha,Kyono:2016jqy,Hoare:2016hwh,Orlando:2016qqu,Borsato:2016ose,Osten:2016dvf,vanTongeren:2016eeb,Hoare:2016wca}, which are defined by $r$-matrices satisfying the homogeneous classical Yang-Baxter equation (cYBE). In turn, the $r$-matrices may be crudely divided into Abelian and non-Abelian, and it has been proved that Abelian $r$-matrices correspond to TsT transformations \cite{Osten:2016dvf}, thus ensuring that the corresponding YB deformation is a supergravity solution. For non-Abelian $r$-matrices, a further ``unimodularity" condition on the $r$-matrix \cite{Borsato:2016ose} distinguishes valid supergravity backgrounds from solutions to so-called generalized supergravity \cite{Arutyunov:2015mqj,Wulff:2016tju}. While a class of non-Abelian $r$-matrices may be understood in terms of non-commuting TsT transformations \cite{Borsato:2016ose}, more generally it has been conjectured (and demonstrated case by case) \cite{Hoare:2016wsk} that homogeneous YB deformations correspond to non-Abelian duality transformations \cite{Fridling:1983ha, Fradkin:1984ai, delaOssa:1992vci, Sfetsos:2010uq, Lozano:2011kb}.
See \cite{Borsato:2016pas} for the proof for the bosonic case
and the result for the supersymmetric case.

This paper constitutes the follow-up to an earlier letter \cite{Araujo:2017jkb}, which made the following key observations and results:

\medskip

\begin{enumerate}
	\item We proved for any YB deformation of AdS$_5$ with an $r$-matrix satisfying the homogeneous cYBE that there is a universal description in terms of open string parameters: concretely, the open string metric~\cite{Seiberg:1999vs} is always the original undeformed AdS$_5$ metric with constant open string coupling\footnote{See \cite{Berman:2000jw} for an earlier observation of this feature in the context of $O(d, d)$ transformations of D-branes.}, and all information about the YB deformation is encoded in the NC structure $\Th$\footnote{To be more accurate, since our $\Theta$ is  not necessarily constant, we call it NC structure, rather than NC parameter which is more usual terminology for constant $\Theta$ cases.}, which in general is an anti-symmetric two tensor all over the AdS$_5 $ bulk.
	\item Based on observation 1. we claimed that homogeneous YB deformations of AdS$_5$ are twists of the conformal algebra $\frak{so}(4,2)$, or alternatively ``conformal twists". Within this setting, this conjecture was made previously in \cite{vanTongeren:2015uha}, where it was argued that YB deformations of AdS$_5$ based on Abelian and Jordanian $r$-matrix solutions to the cYBE result in Drinfeld twists of the conformal algebra. Where the Drinfeld twists of the conformal algebra have been studied, namely for the Poincar\'e subalgebra \cite{Chaichian:2004za, Chaichian:2004yh, Lukierski:2005fc}, one can confirm that the NC structure $\Theta^{\mu \nu}$, defined through the characteristic commutation relation involving spacetime operators $\hat{x}^{\mu}$,
	\be
	\label{NC_spacetime}
	\left[ \hat{x}^{\m}, \hat{x}^{\n} \right] = i \Th^{\m \n} \qquad (\m,\n=0,\ldots,3),
	\ee
	agrees with the NC structure extracted from the geometry \textit{at leading order}. Working at this order, we identify a large class of NC structures from the geometry, as well as recovering examples based on well-defined Drinfeld twists of Poincar\'e algebra. Since a twist is expected for each $r$-matrix, although the precise form of the twist element may be unknown, one can expect agreement at leading order.
	\item Our almost exhaustive list of examples led us to the novel concept of ``holographic noncommutativity". In extending from Poincar\'e twists to twists of the full conformal algebra, one may consider twists involving dilatation $D$ and special conformal generators $K_{\mu}$, which result in an NC structure dependent on the AdS radius $z$. Since the generators of the conformal algebra are Killing vectors in the bulk, with the Killing vectors corresponding to $D$ and $K_{\mu}$ picking up a $z$ dependence, it is most natural to consider this holographic description. The holographic (bulk) NC structure $\Theta^{MN}$, $M, N = 0, \dots, 3, z$ and the boundary NC structure $\Theta^{\mu\nu}$ (\ref{NC_spacetime}) are related as
	\be\label{holographic-NC}
	\Theta^{\mu z}(z=0)=0,\qquad \Theta^{MN}(z=0)=\Theta^{\mu\nu}.
	\ee
	Moreover, the value of $\Theta^{MN}$ everywhere in the bulk is uniquely specified by the value of $\Theta^{\mu\nu}$ at the boundary, thereby justifying the ``holographic noncommutativity'' picture.

%The fact that bulk and boundary NC structures are in general different is an important distinction to earlier works \cite{vanTongeren:2015uha, vanTongeren:2016eeb}, where only Poincar\'e twists are considered. For sample conformal twists, we will later demonstrate that the geometric NC structures, evaluated at the boundary, agree with the NC structures defined through the Drinfeld twist. This provides further evidence for the conjecture.
	
	\item
	Based on exhaustive case by case studies, we show that the NC structure read off from the geometry is directly related to the $r$-matrix
	through the deformation parameter $\eta$,
	\be\label{Theta-r}
	\Th^{MN} = -2 \, \h \,{\hat{r}^{MN}} \qquad (M,N=0,\ldots,3,z),
	\ee
where $\hat{r}^{MN}$ denote the components of the $r$-matrix evaluated in the standard coordinate basis vectors. It is worth noting that the above relation (\ref{Theta-r}) is holographic in nature and setting $z=0$ one recovers an equation that was first recorded in \cite{vanTongeren:2016eeb} for a large class of unimodular models. However, as we argue in section \ref{sec:open_string}, it is more general and holds for all homogeneous YB deformations. In fact this direct relation between $\Theta$ and the $r$-matrix is even more general. To illustrate this fact, in appendix \ref{sec:mCYBE} we show a direct relation between $\Theta$ and the tensorial $r$-matrix also exists for YB deformations based on $r$-matrix solutions to the modified cYBE \cite{Delduc:2013qra}.

	\item For YB deformations corresponding to valid supergravity solutions, i. e. unimodular $r$-matrices, we invoked AdS/CFT to infer the existence of corresponding NC deformations of $\mathcal{N}=4$ sYM with NC structure $\Theta^{\mu \nu} = - 2 \, \eta \, {\hat{r}^{\mu \nu}}$. We stress that similar conclusions have been arrived at by van Tongeren in \cite{vanTongeren:2015uha, vanTongeren:2016eeb} and here we provide a complementary viewpoint. We recall that following analysis pioneered by Maldacena-Russo \cite{Maldacena:1999mh} in the AdS/CFT context, van Tongeren considered deformations of D3-brane geometries and the low-energy decoupling limit. Here, we work directly in the near-horizon limit, and instead of discussing the decoupling limit, we invoke now standard AdS/CFT logic that any deformation in the gravity side should correspond to a deformation in the sYM side and based on global symmetries preserved by the deformations, we   motivate our conjectured duality to NC deformations of super Yang-Mills.\footnote{For Drinfeld twists based on Poincar\'e symmetries, which correspond to TsT transformations, the decoupling limit of Dp-branes and the map between closed and open string frames trivially commute. This is clear from the analysis presented in section \ref{sec:open_string}, where we show that for a simple TsT and a general spacetime, the NC structure is a constant, so for Dp-branes it is the same before and after the decoupling limit.}  For twists based on the conformal group, not only do our mathematical relations and general AdS/CFT expectations support existence of dual NC sYM description, but also as we argue, there is an automorphism of the conformal algebra which relates these twists to the Poincar\'e twists. This latter provides another justification for the presence of such a decoupling limit.

	\item Non-unimodular YB deformations result in solutions of generalized supergravity, which are specified by a Killing vector field $I$ \cite{Arutyunov:2015mqj,Wulff:2016tju}: simply put, setting $I = 0$, we recover usual supergravity. We identified a novel relation that bridges the open string and closed string descriptions,
	\be\label{unimodular}
	\nabla_{M} \Th^{MN} = I^{N}.
	\ee
	As a direct consequence, unimodular YB deformations are characterized by divergence-free $\Theta^{MN}$. For these cases, the divergence-free condition allows solving $\Theta^{MN}$ as a bulk field in terms of the boundary NC structure $\Theta^{\mu\nu}$, thus justifying the holographic noncommutativity discussed in item 2. above. We attributed this remarkable result to the preservation of $\Lambda$-symmetry \cite{Witten:1995im,SheikhJabbari:1999ba}, thus providing the first explanation of the unimodularity condition \cite{Borsato:2016ose} where the Killing vector field $I$ is directly related to the $r$-matrix through the NC structure.  %\footnote{\red{Strictly speaking, the unimodularity condition can be derived from Weyl invariance of the $\sigma$-model. A remarkable point about equation (\ref{unimodular}) is that it is still valid
%beyond the on-shell condition of the usual supergravity, turning upon non-vanishing $I$.}
%outside of the supergravity regime,
%where the right hand side gives information about the Killing vector $I$.
%}.
	\end{enumerate}

\medskip

In this paper, we extend the results of \cite{Araujo:2017jkb} in the following ways. Firstly, we present a proof of the relation (\ref{Theta-r}) for AdS$_5$ spacetimes with Poincar\'e metric. In addition, we discuss $\Lambda$-symmetry of D-branes and provide the details of how it accounts for the equation (\ref{unimodular}). Along the way, we provide a further check of the consistency of the equations of motion of generalized supergravity \cite{Arutyunov:2015mqj,Wulff:2016tju} (see appendix \ref{sec:gen_IIB}). We further discuss equation (\ref{unimodular}) in terms of a candidate Poisson structure defined by the NC structure $\Theta$ and we show that the Killing vector $I$, pulled back to the D-brane worldvolume, describes a modular vector field corresponding to a candidate Poisson structure defined by $\Theta$ (see appendix \ref{appen:cohomology}).  As extra checks on our results, we identify $\Theta$ and $I$ for a large class of YB deformations, thus allowing  the reader the opportunity to confirm the validity of equation (\ref{unimodular}) directly.

Some of the examples of YB deformations we discuss are new to the literature. The motivation to study them stems from the fact that the outer automorphism of the conformal algebra, $D \leftrightarrow -D$, $P_{\mu} \leftrightarrow K_{\mu}$, which flips the sign of the dilatation operator, while interchanging translation and special conformal symmetries, corresponds to a coordinate transformation in anti-de Sitter. We explicitly check that YB deformations corresponding to $r$-matrices related through this automorphism, are in turn related geometrically via coordinate transformation. As a second application, we demonstrate that the maximally symmetric AdS$_p\times$S$^p$ geometries, $p =2, 3, 5$, are self-T-dual under combination of bosonic T-dualities \cite{Buscher:1987sk, Buscher:1987qj} along \textit{special conformal isometries} and fermionic T-dualities \cite{Berkovits:2008ic, Beisert:2008iq} with respect to \textit{isometries constructed from superconformal Killing spinors}. This should be contrasted with the usual prescription where bosonic T-dualities along translation isometries and fermionic T-dualities with respect to Poincar\'e Killing spinors feature \cite{Berkovits:2008ic, OColgain:2012ca} (see also \cite{Adam:2009kt, Dekel:2011qw}).

The structure of this paper is as follows. Following a brief review of YB deformations in section \ref{sec:YBreview},  in section \ref{sec:open_string} we introduce and define open string parameters and prove that the open string metric and coupling constant agree with the undeformed closed string metric and coupling constant.  In section \ref{sec:twists}, we invoke AdS/CFT to make the case for unimodular YB deformations being dual to noncommutative deformations of sYM. In section \ref{sec:unimodular}, we review the $\Lambda$-symmetry of D-branes and  explain how it accounts for equation (\ref{unimodular}). In section \ref{sec:examples}, we provide a host of examples of YB deformations, some of which are new. In each case, we provide explicit expressions for $\Theta$ and $I$, allowing the reader to validate (\ref{unimodular}) for a large class of examples. Finally, in section \ref{sec:auto} we introduce a known outer automorphism of the conformal algebra that corresponds to a coordinate change in AdS. We show that the coordinate transformation may be used to generate YB deformations, as well as illustrating self-T-duality based on a combination of T-dualities with respect to special conformal symmetries and superconformal supercharges. Section \ref{discussion-section} contains our concluding remarks. In the four appendices we have gathered part of our conventions and some further technical details of the computations and analysis.

\section{Review of the YB deformation}
\label{sec:YBreview}
We begin with a review of the YB deformed $\sigma$-model following the coset construction. Our presentation is a simplification of the AdS$_5\times$S$^5$ supercoset construction presented in  \cite{Kyono:2016jqy} where, since we are primarily interested in deformations of AdS$_5$, we consider the coset space $SO(4,2)/SO(4,1)$ and neglect the internal space. Moreover, we restrict our attention to YB deformations based on $r$-matrix solutions to the homogenous cYBE.

The YB deformed $\sigma$-model action is \cite{Matsumoto:2015jja,Kawaguchi:2014qwa, Delduc:2013qra, Delduc:2013fga},
\be
\label{YB_deformed}
\mathcal{L} = \textrm{Tr} \left[ A\,P^{(2)} \circ \frac{1}{1 - 2 \h R_g \circ P^{(2)}} A\right],
\ee
with a deformation parameter $\h$ and $R_{g} (X) \equiv g^{-1} R(g X g^{-1}) g$\,. Setting $\h = 0$, we recover the undeformed AdS$_5$ $\sigma$-model.
Above, $A=-g^{-1}\dd g$, is a left-invariant current with $g$ an element of the conformal group, $g \in SO(4,2)$. Moreover,
$P^{(2)}$ is a projector onto the coset space $\mathfrak{so}(4,2)/\mathfrak{so}(4,1)$, spanned by the generators ${\bf P}_{m}\,(m=0,\ldots,4)$, which satisfy $\textrm{Tr}[{\bf P}_{m}{\bf P}_{n}]=\h_{mn} =\textrm{diag}(-++++)$. $P^{(2)}$ may be expressed as
\be
P^{(2)}(X)=\h^{mn}{\rm Tr}[X\,{\bf P}_{m}]\,{\bf P}_{n}\,,\quad X\in \frak{so}(4,2)\,.
\ee
We summarize our matrix representations in appendix \ref{sec:conventions}. For further details, we refer the reader to  \cite{Kyono:2016jqy}.

Returning to the exposition of the deformation, $R$ is an antisymmetric operator satisfying the homogeneous cYBE,
\be
\label{cYBE}
[ R (X), R (Y)] - R([R (X), Y]+ [X, R (Y)])=0,
\ee
with $X, Y \in \frak{so}(4,2)$.
In turn, the operator $R$ can be written in terms of an $r$-matrix as,
\be
R(X)  =\textrm{Tr}_2[r(1\otimes X)] =\sum_{i,j}r^{ij} b_i \textrm{Tr} [ b_j X],
\ee
where $r\in \frak{so}(4,2)\otimes \frak{so}(4,2)$ is
\be
\label{r-matrix}
r =\frac12 \sum_{i,j}  r^{ij} b_i\wedge b_j, \quad {\rm with}\quad b_i\in \frak{so}(4, 2).
\ee
The $r$-matrix is called Abelian if $[b_i,b_j]=0$ and unimodular if
it satisfies the following condition \cite{Borsato:2016ose}:
\be\label{unimodular_r}
r^{ij}[b_i,b_j]=0,
\ee
where $i, j$ range over the generators of $\frak{so}(4,2)$.

%but it should not be confused with $r^{MN}$ appearing in (\ref{Theta-r}), as this is the $r$-matrix expressed differential operators on AdS$_5$.
%This distinction will be important when we come to discuss holographic noncommutativity.

In order to read off the YB deformed geometry from (\ref{YB_deformed}), we adopt the following parametrization for $g$,
\be
g = \exp [ x^{\m} P_{\m} ] \exp [ (\log z) D],
\ee
where $P_{\m}\,(\m=0,...,3)$, $D$ respectively denote translation and dilatation generators, which are related to ${\bf P}_{m}$ as explained in the appendix. Having introduced coordinates, we are now in a position to define $\hat{r}^{MN}$ as
\be
\hat{r} = \frac{1}{2} \hat{r}^{MN} \partial_{M} \wedge \partial_{N},
\ee
where $\partial_{M}$ are differential operators on AdS$_5$ (\ref{diff_operators}) with $M, N$ ranging over the holographic coordinate, namely $z$.
Then,
%\red{ defining the dressed $r$-matrix generators $a_{i \, g} \equiv g^{-1} a_i g = a_{i\, g}^{\m} P_{\m} + a_{i \,g}^5 D$,}
the YB deformed metric $g_{MN}~(M,N=0,\ldots,3, z)$, NS-NS two-form $B_{MN}$, and dilaton $\Phi$  (in string frame) can be expressed as \cite{Kyono:2016jqy},
\be\label{YB_deformation}
\bg
g_{MN} = e^{m}_{M} e^{n}_{ N} k_{(mn)},\quad B_{MN} = e^{m}_{M} e^{n}_{N} k_{[nm]}, \quad {\rm e}^\Phi = (\textrm{det}_5\, k)^{1/2},\\ k_{mn}  = k_{(mn)} + k_{[mn]},
%( 1+ 2 \h \lambda),
\eg\ee
where $e^{m}_{M}$ is the AdS$_5$ vielbein, and we have defined,
\bsub\begin{align}
\label{k} k_{m}{}^{n}&\equiv (  \d^{m}{}_{n} - 2 \h \lambda^{m}{}_{n} )^{-1}, \\
\label{lambda} \lambda_{m}{}^{n} &\equiv
\h^{nl}{\rm Tr}[{\bf P}_{l}R_g({\bf P}_{m})].
\end{align}\esub
This completes our review of YB deformations.

To close this section, it is instructive to consider an example. We return to the earliest example of an Abelian twist \cite{Hashimoto:1999ut,Maldacena:1999mh}, which will serve as our most pedestrian example in this work and  in this case we will be deliberate. Here, the Abelian $r$-matrix  \cite{Matsumoto:2014gwa} is,
\be
r = \frac12 P_1 \wedge P_2.
\label{PP}
\ee
In terms of differential operators, the $r$-matrix may be expressed as
\be
{\hat{r} = \frac12 \partial_1 \wedge \partial_2. }
\label{PP_diff}
\ee

To determine the deformation (\ref{YB_deformation}), we must first identify $\lambda_{m}^{~n}$ (\ref{lambda}), before plugging back into (\ref{k}). We initially determine (see appendix \ref{sec:conventions}),
\bea
g^{-1} P_{i} g = \frac{1}{z} P_i, \quad R_{g} ( {\bf P}_{1}) = \frac{1}{2 \, z^2} P_2, \quad R_{g} ( {\bf P}_{2}) = -\frac{1}{2 \, z^2} P_1.
\eea
Next, we substitute this expression into (\ref{lambda}) to identify the only non-zero component of $\lambda_{m}^{~n}$, namely
\bea
\lambda_{1}^{~2} = \textrm{Tr} [ {\bf P_2} R_{g} ({\bf P}_{1})] = -  \frac{1}{2 \, z^2}, \quad  \lambda_{2}^{~1} = \textrm{Tr} [ {\bf P_1} R_{g} ({\bf P}_{2})] = \frac{1}{2 \, z^2} .
\eea

The NSNS sector of the corresponding closed string solution may then be read off from (\ref{YB_deformation}),
\be\bn
 \dd s^2 &=  \frac{(- \dd t^2 + \dd x_3^2 + \dd z^2)}{z^2}
+ \frac{z^2}{(z^4 + \h^2)} (\dd x_1^2 + \dd x_2^2)+ \dd s^2(S^5) , \\
 B  &=  \frac{\h }{z^4 + \h^2} \dd x_1 \wedge \dd x_2, \quad
e^{2 \Phi} =g^2_s \frac{z^4}{(z^4 + \h^2)}.
\label{MR}
\en\ee
where we have omitted the RR sector, which may be found in the original texts \cite{Hashimoto:1999ut,Maldacena:1999mh}. For the full supercoset construction of this background, see \cite{Kyono:2016jqy}. This solution was initially generated using a TsT transformation starting from the AdS$_5\times$S$^5$ solution to type IIB supergravity.

\section{``Open string'' description}
%\red{Add comments on open strings. See comment 2. of report}
\label{sec:open_string}
Having introduced the nuts and bolts of the YB deformation, although admittedly in an abstract fashion, in this section we outline one of the novel results of our work. The viewpoint we adopt here is that  YB deformations of AdS$_5\times$S$^5$ $\sigma$-model,  can be viewed as field redefinitions of supergravity fields defining the background. This field redefinition, for historical reasons related to the context they were first coined and used, is called closed string to open string map. That is, for a given deformed background supergravity solution specified by ``closed string'' metric, $B$-field and dilaton,  one can define the corresponding ``open string'' metric, $\Theta^{MN}$-field and open string coupling. As we will prove below, it turns out that the open string  is the original \textit{undeformed} metric with \textit{constant} string coupling. In this description, the field redefinition/deformation is hence fully parameterized only by the so-called NC structure $\Theta^{MN}$.  We note that although we call $\Theta^{MN}$ the NC structure it is in general an antisymmetric two tensor field defined in the bulk.

To be more concrete, we introduce closed string parameters ($g_{MN}$, $ B_{MN}$, $g_s$), which allow us to define the open string metric $G_{MN}$, NC structure $\Th^{MN}$ and coupling $G_s$ \cite{Seiberg:1999vs}:
\begin{align}
G_{MN} &= \left( g- Bg^{-1}B\right)_{MN},\label{Open-metric} \\
\Th^{MN} &= - \left( ({g + B})^{-1} B ({g - B})^{-1} \right)^{MN},\label{Theta-def} \\
G_s &= g_s {\rm e}^{\Phi} \left( \frac{\det (g+B)}{ \det g} \right)^{\frac{1}{2}} \label{Gs}.
\end{align}
We stress that the above, and at this stage, should be viewed as a field redefinition from the so-called ``closed-string frame'' $(g_{MN}, B_{MN}, \Phi)$ to the ``open string frame'' $(G_{MN}, \Theta^{MN}, G_s)$.\footnote{The context in which the names open string and closed string frame were  first coined \cite{Seiberg:1999vs} and find their precise meaning, is a flat space with constant $B$-field. Nonetheless, one can show that this field redefinition appears for general backgrounds in the context of T-duality as canonical  transformation from a worldsheet viewpoint, e.g. see \cite{Open-closed}.} As we will argue, however, the name open string frame is very well justified since fields in this frame describe the open string degrees of freedom, the massless sector of which is associated with the dual NC sYM description residing on the boundary of the (deformed) AdS space.

Before progressing to general homogeneous YB deformations of AdS$_5$, which is the focus of this paper, we meander somewhat to discuss deformations based on Abelian $r$-matrices. It has been proved that YB deformations with Abelian $r$-matrices correspond to TsT transformations of superstring $\sigma$-models \cite{Osten:2016dvf}. So, as a warm-up, it is prudent to check for TsT transformations that the open string metric is undeformed. \footnote{It is well-known that the effect of TsT transformations for the AdS$_5\times$S$^5$ background can be undone by mapping the deformed closed string $\sigma$-model action to the action for open strings in AdS$_5\times$S$^5$ with twisted boundary conditions for the open strings \cite{Frolov:2005dj, Frolov:2005iq, Alday:2005ww}, yet in these works the connection to the Seiberg-Witten closed-string to open-string map \cite{Seiberg:1999vs} is not made. This motivates us to make this connection explicit and, in the process, comment on how the NS-NS two-form affects the analysis.}

To this end, we consider the ansatz for the NS sector of 10D supergravity with $U(1) \times U(1)$ isometry \cite{Colgain:2016gdj},
\bea
\dd s^2_{10}  &=& \dd s^2_8 + e^{2 C_1} \mathcal{D} \varphi_1^2  + e^{2 C_2} \mathcal{D} \varphi_2^2, \label{TsT_original_metric} \\
B &=& B_2 + \mathcal{B}^1 \wedge \dd \varphi_1 + \mathcal{B}^2 \wedge \dd \varphi_2,
\eea
where we have defined the covariant derivatives $\mathcal{D}  \varphi_i = \dd \varphi_i + \mathcal{A}^i$. Here, $C_i$ denote scalar warp factors and $\mathcal{A}^i, \mathcal{B}^i$ represent gauge fields that only depend on the transverse 8D spacetime. Neglecting cross-terms in the metric and $B$-field , i. e. $g_{\varphi_1 \varphi_2}, B_{\varphi_1, \varphi_2}$, this is the most general ansatz. We also allow for a non-trivial dilaton $\Phi$.

Performing T-duality on $\varphi_1$, a constant shift $\varphi_2 \rightarrow \varphi_2 + \eta \varphi_1$, and a second T-duality back along $\varphi_1$, we find the TsT-deformed {NS-NS} sector  \cite{Colgain:2016gdj}:
\be\bn
\label{Tdual_NS}
\dd \tilde{s}^2_{10} &= \dd s^2_8 + \frac{1}{[ 1 + \h^2 e^{2 C_1 + 2 C_2} ]} \biggl[ e^{2 C_1} ( \mathcal{D} \varphi_1 + \eta \mathcal{B}^2)^2 + e^{2 C_2} (\mathcal{D} \varphi_2 - \eta \mathcal{B}^1)^2 \biggr], \\
\tilde{B} &= B_2 + \mathcal{B}^1 \wedge \dd \varphi_1 + \mathcal{B}^2 \wedge \dd \varphi_2 + \h \mathcal{B}^1 \wedge \mathcal{B}^2 \\
&- \frac{\lambda e^{2 C_1 + 2 C_2}}{[ 1 + \h^2 e^{2 C_1 + 2 C_2}]} (\mathcal{D} \varphi_1 + \h \mathcal{B}^2) \wedge (\mathcal{D} \varphi_2 - \h \mathcal{B}^1), \\
\tilde{\Phi} &= \Phi - \frac{1}{2} \ln ( 1+ \h^2 e^{2 C_1 + 2 C_2}).
\en\ee
Using the above expression, it is easy to check that the open string metric (\ref{Open-metric}) is simply the original metric (\ref{TsT_original_metric}) on the condition that $B_2 = \mathcal{B}^i = 0$, so that initially there can be no NS-NS two-form $B$. Setting $B=0$, we find that we recover the original string coupling, $G_s = g_s$, while the NC structure is simply a constant $\Theta^{\varphi_1 \varphi_2} = \eta$. It is worth stressing that the result is not completely general. We must ensure that there is no NS-NS two-form before we start the deformation. In the existing YB deformation literature, since most works focus on deformations of AdS$_5\times$S$^5$, where $B=0$, this is tacitly assumed. \footnote{As a side remark, we note that the geometry AdS$_3 \times$S$^3$ can be supported purely by an {NS-NS} three-form flux, where the action of T-duality is obstructed by the flux threading AdS$_3$ and S$^3$. This would appear to preclude TsT transformations, so it may be interesting to study YB deformations in that context to understand if there is an issue, potentially some breakdown in the YB machinery reviewed in section \ref{sec:YBreview}. }

Next, we specialize to the cosets, where we discuss deformations based on non-Abelian $r$-matrices. More concretely, for YB deformations of AdS$_5$ \eqref{YB_deformation}, it is easy to identify the inverse open string metric and NC structure,
\be\label{3.7}
G^{MN} + \Th^{MN} = e^{M}_{m} e^{N}_{n} \left( \h^{mn} + 2 \h \,
\lambda^{mn} \right),
\ee
where $e^{M}_{m}$ denotes the inverse vielbein. Note, the classical Yang-Baxter equation is not needed to derive~\eqref{3.7}. Since $\lambda^{mn}$ is anti-symmetric, it is straightforward to separate the components,
\be
\label{G_theta}
G^{MN} = e^{M}_{m} e^{N}_{n} \h^{mn}, \quad \Th^{MN} =  2 \h \,e^{M}_{m} e^{N}_{n} \, \lambda^{mn}.
\ee
Inverting $G^{MN}$,  it is clear that the open string metric is precisely the original AdS$_5$ metric. Moreover, inserting \eqref{YB_deformation} into \eqref{Gs}, we get $G_s=g_s = {\rm constant}$. That is, all the information about the YB deformation, as viewed by open strings, is sitting in $\Th^{MN}$, while the metric is undeformed AdS$_5$.

We now return again to our familiar example (\ref{MR}), where the open string metric becomes,
\bea
\dd s^2_{\textrm{open}} &=& \frac{1}{z^2} ( - \dd x_0^2 + \dd x_1^2  + \dd x_2^2 + \dd x_3^2 + \dd z^2) + \dd s^2(S^5), \nn
\Theta^{12} &=& - \eta, \quad G_s = g_s.
\eea
We recognize immediately that the equation (\ref{Theta-r}) is satisfied. It is also worth noting that since the NC structure is constant, the divergence trivially vanishes and we see that the geometry is a solution to supergravity, as expected.  Furthermore, while the closed string metric \eqref{MR} has a severely deformed causal and boundary structure \cite{Hashimoto:1999ut, Maldacena:1999mh, Alishahiha:1999ci}, the spacetime as seen by the open strings is the usual \adss with $\mathbb{R}^{1,3}$ boundary. The open string parameters (in particular the metric) constitute the relevant geometry for the dual field theory and therefore our result above indicates that the dual  description is a $\Theta$-deformed sYM.

As a final comment, let us note that $G_s = g_s$ implies the determinant of the open string and closed string metrics are related as follows (see (2.44) of \cite{Seiberg:1999vs}):
\be
\sqrt{G} = e^{-2 \Phi} \sqrt{g}.
\ee
Since the open string metric is undeformed, this implies that the density $e^{-2 \Phi} \sqrt{g}$, in addition to being a recognized invariant of T-duality, is also an invariant of YB deformations.

\subsection{Proof of $\Theta^{MN} = - 2 \, \eta \, \hat{r}^{MN}$}
Here, we sketch a proof of equation (\ref{Theta-r}) for YB deformations of AdS$_5$. To begin, we note that $M, N$ range over the holographic direction $z$ in addition to the Minkowski coordinates, $\mu = 0, \dots, 3$. Given an $r$-matrix (\ref{r-matrix}), $\lambda^{mn}$ may be re-expressed as,
\be
\lambda^{mn} = \sum_{i,j} \eta^{mk} \eta^{nl} r^{ij} \textrm{Tr} [ {\bf P}_l (g^{-1} b_i g) ] \textrm{Tr} [ {\bf  P}_{k} (g^{-1} b_j  g)],
\ee
where $m, n = 0, \dots, 4$ range over the generators of the coset, while $i, j$ enumerate generators in the $r$-matrix. We hope there is no confusion.

Therefore to evaluate (\ref{G_theta}), we require only to evaluate $ \textrm{Tr} [ {\bf P}_{l} (g^{-1} b_i g) ] $ for the generators of the conformal algebra, $b_i \in \{ P_{\mu}, M_{\mu \nu}, D, K_{\mu} \}$. Using the matrix representation in the appendix \ref{sec:conventions}, the non-zero expressions are:
\bea
\textrm{Tr} [ {\bf P}_{l} (g^{-1} P_{\mu} g) ]  &=& - \frac{1}{z} \, \eta_{l \mu}, \nn
\textrm{Tr} [ {\bf P}_{l} (g^{-1} M_{\mu \nu} g) ]  &=& - \frac{x_{\nu}}{z} \, \eta_{l \mu} + \frac{x_{\mu}}{z} \, \eta_{l \nu}, \nn
\textrm{Tr} [ {\bf P}_{l} (g^{-1} D g) ] &=& - \frac{x_{l}}{z}, \quad \textrm{Tr} [ {\bf P}_{4} (g^{-1} D g) ] = -1, \nn
\textrm{Tr} [ {\bf P}_{l} (g^{-1} K_{\mu} g) ]  &=& \frac{2 x_{l} x_{\mu}}{z}  - \frac{ x_{\nu} x^{\nu} + z^2}{z} \eta_{l \mu}, \quad  \textrm{Tr} [ {\bf P}_{4} (g^{-1} K_{\mu} g) ] = 2 x_{\mu}.
\eea
The important point to note is that the expressions simply correspond to the components of the same generator evaluated as a differential operator (\ref{diff_operators}) once the $z$-dependence is ignored. This makes it easy to simply determine $\lambda^{mn}$ and substitute it back in (\ref{G_theta}), where factors of the inverse vielbein of AdS$_5$ ensure that the correct $z$ factors appear in the final expression for the differential operators.

 \section{Conformal Twists \& NC gauge theory}
 \label{sec:twists}

Quantum field theory (QFT) can be formulated on the NC spacetime characterized by~$\Theta$~\eqref{NC_spacetime}. To illustrate how this comes about, first consider the case of constant $\Theta$, which arises in particular in the example~\eqref{MR}. One can replace the product of spacetime fields in the QFT with the Moyal star product in the following way,
\be
\label{moyal_star}
f(x) g(x) \rightarrow (f \star g) (x) = f(x) {\rm e}^{\frac{i}{2} \Theta^{\mu \nu} \overset{\leftarrow}{\partial_{\mu}} \overset{\rightarrow}{{\partial}_{\nu}} } g(x).
\ee
The commutator then turns into the Moyal bracket:
\be
\label{moyal_bracket}
[f, g]_{\star} := f \star g - g \star f = i \Theta^{\mu \nu} \partial_{\mu} f \partial_{\nu} g + \mathcal{O} ( \partial^3 f, \partial^3 g).
\ee
It should be noted that we recover the commutator~\eqref{NC_spacetime} when $f(x) = x^{\mu}, g(x) = x^{\nu}$.

It is known that a theory with Moyal $\star$-product is equivalent to employing a Hopf algebra with the co-product that uses an Abelian Drinfeld twist element~\cite{Chaichian:2004za},
\be
\label{twist}
\mathcal{F} = {\rm e}^{-2i\eta r} = {\rm e}^{ \frac{i}{2} \Theta^{\mu \nu} P_{\mu} \wedge P_{\nu} },
\ee
where the $r$-matrix appearing here satisfies the cYBE~\cite{Matsumoto:2014gwa}. By replacing the matrix generators $P_{\mu}$ with differential operators $\hat{P}_{\mu}$, the Moyal star product (\ref{moyal_star}) can be reconstructed. The twist~\eqref{twist} is Abelian and it does not change the Poincar\'e algebra~\cite{Chaichian:2004za}, instead the co-product of the universal enveloping algebra of the Poincar\'e algebra is deformed~\cite{Drinfeld}.

For the simplest case of the twist~\eqref{twist}, it turns out that the resulting NC structure is constant. In fact, for any given twist, the NC structure is determined: if one changes the $r$-matrix appearing in the twist, this change is reflected in the NC structure of the spacetime. Going beyond the simple Abelian twist above, if one allows for Lorentz generators $M_{\mu\nu}$, one can consider solutions to the cYBE of the form $r \sim P \wedge M$ and $r \sim M \wedge M$. These correspond to resulting NC structures that are linear and quadratic in Cartesian coordinates, respectively~\cite{Lukierski:2005fc}. As a concrete example, consider replacing, for example, $r = P_1 \wedge P_2$ with $r = \frac{1}{2} M_{01} \wedge M_{23}$ in the twist~\eqref{twist}. Then up to a sign that depends on conventions, $\Theta$ is given by~\cite{Lukierski:2005fc}:
\be\begin{split}
\Theta^{02} =& - 2 \sinh \frac{\eta}{2} \cdot x^1 x^3, \quad
 \Theta^{03} =  2 \sinh \frac{\eta}{2} \cdot x^1 x^2, \\
\Theta^{12} =& - 2 \sinh \frac{\eta}{2} \cdot x^0 x^3, \quad
\Theta^{13} =  2 \sinh \frac{\eta}{2} \cdot x^0 x^2.
\end{split}
\ee
We have checked that the YB prescription~\eqref{G_theta} can be used to obtain the same expressions at leading order in $\eta$. On the basis of this example, we are forced to conclude that our method for reading off $\Theta$ from the geometry is only sensitive to terms linear in the deformation parameter.

We have exhausted all known twists of the Poincar\'e subalgebra of the conformal algebra. To the extent of our knowledge at this point the literature terminates, so we are unable to compare further\footnote{More generally, one may consider $\kappa$-Minkowski spacetime and the associated $\kappa$-Poincar\'e algebra, but it is worth noting that it cannot be obtained from a twist of the usual Poincar\'e algebra \cite{Borowiec:2013gca, Dimitrijevic:2014dxa}.}. However, we conjecture that homogeneous YB deformations of anti-de Sitter correspond to twists of the conformal algebra, so that the known results from YB deformations may serve as predictions for the corresponding conformal twists. We will now take a closer look at two examples.

Let us consider the $r$-matrices,
\be
\begin{split}
\label{r1,2} r_1 =& \frac{1}{2}D \wedge K_1, \\
 r_2 =& \frac{1}{2}  (P_0 - P_3) \wedge (D + M_{03}).
\end{split}
\ee
Since they involve dilatation $D$ and special conformal generators $K_{\mu}$, we regard these as conformal twists in the true sense. We note that the latter is unimodular, in fact Abelian\footnote{
It is worth noting the following relation
\[
[P_0-P_3, D + a M_{03}] = (a-1)(P_0-P_3),
\]
where $a$ is a constant. Hence $r_2$ in (\ref{r1,2}) is Abelian because it corresponds to the case with $a=1$.
} (e.g., see the classification of~\cite{Borsato:2016ose}), while the former is not unimodular, so that the corresponding YB deformation does not lead to a supergravity solution. Making use of ~\eqref{lambda} and~\eqref{G_theta} it is straightforward to find the values of $\Theta$:
\bsub\be\label{Theta-r1}
\Theta_{(1)}^{1 {\mu}} = \eta x^{{\mu}} ( x_{\nu} x^{\nu}  + z^2), \qquad \Theta_{(1)}^{1z} = \eta z ( x_{\nu} x^{\nu}  + z^2), \ee
\be\label{Theta-r2}
\Theta_{(2)}^{-+} = -4 \eta x^+, \quad\Theta_{(2)}^{-i} = -2 \eta x^i, \quad\Theta_{(2)}^{- z} = -2 \eta z,
\ee\esub
where $\mu \neq1$, $i = 1, 2$ and $x^{\pm} = x^0 \pm x^3$. $\Theta_{(1,2)}$ correspond to the $r$-matrices $r_{1,2}$ respectively. One can readily confirm that~\eqref{Theta-r} is satisfied. From the explicit form of the conformal algebra generators~\eqref{diff_operators} it is obvious that in general the NC structure may have cubic and quartic terms.

Note that the NC structure may have non-trivial $z$-dependence with non-zero legs $\Theta^{z\mu}$ along $z$. However, as is explicitly seen for the examples in \eqref{Theta-r1}, \eqref{Theta-r2}, all the $\Theta^{z\mu}$ components vanish at the AdS boundary at $z=0$. Referring the reader to the explicit expressions for the generators of conformal algebra as differential operators (\ref{diff_operators}), it is clear that at $z = 0$, we simply recover the standard form of the four dimensional conformal algebra generators. Moreover, the value and form of $\Theta^{MN}$ components at arbitrary $z$ in the bulk is determined from $\Theta^{\mu\nu}$ at the boundary. In this sense all the information stored in $\Theta^{MN}$, including its $z$-dependence, can be recovered from the boundary information.

Unimodular $r$-matrices lead to the YB deformations that have an interpretation as deformed AdS$_5$ string theory backgrounds. Within AdS/CFT logic, closed string theory in the decoupling limit on these backgrounds is expected to be dual to Yang-Mills theories with noncommutativity parameter $\Theta^{\mu \nu} = -2\,\eta\, \hat{r}^{\mu\nu}$~\cite{Hashimoto:1999ut, Maldacena:1999mh, Alishahiha:1999ci}. Note, however, that the existence of a decoupling limit should not be taken for granted~\cite{Alishahiha:1999ci} (see also \cite{vanTongeren:2015uha, vanTongeren:2016eeb} for related discussion). It has been argued that in the case of ``electric'' noncommutativity ($\Theta^{\mu\nu} \Theta_{\mu\nu} < 0$) the open string theory does not reduce to NC sYM. One then has to deal with the non-critical NC open string theory (NCOS) \cite{Seiberg:2000ms,Ganor:2000my,Gopakumar:2000na,Russo:2000zb}, which is related to  NC sYM at strong coupling. While the analysis of the mentioned papers and most of the literature on NC field theories are made for the Moyal case with constant $\Theta$, it is plausible that these results extend over to generic $x$ dependent NC cases.

As an important final comment in this section, we would like to point out that as discussed in \cite{Beisert:2005if}, the planar integrability of ${\cal N}=4$ sYM continues to hold for a certain subset of Abelian twists of the Poincar\'e algebra. Planar integrability of the NC sYM and analysis of \cite{Beisert:2005if} is established as a perturbative result in the small 't Hooft coupling, and recalling the integrability results for undeformed sYM, it is presumably extendable to all orders in 't Hooft coupling. The integrability of the open string (or closed string) worldsheet theory, however, via the AdS/CFT, is translated to the planar integrability at strong 't Hooft coupling. The two integrabilities hence confirm and strengthen each other. In the NC case, the integrability of the worldsheet along with analysis of \cite{Beisert:2005if}, may be used to conjecture the planar integrability of the  dual NC sYM. See \cite{vanTongeren:2015uha} for earlier comments in this direction. We shall return to this point later in the discussion section.

\subsection{Comparison to Drinfeld twists}
In this subsection, we will show that NC structures read off from the open string description, namely (\ref{Theta-r1}) and (\ref{Theta-r2}), agree with the corresponding NC structures that arise from Drinfeld twists of the conformal algebra. To the extent of our knowledge, these results on twists of the conformal algebra are new and help substantiate our claim that homogeneous YB deformations of unimodular $r$-matrices are dual to conformal twists of Yang-Mills theories. In fact, as the astute reader will notice, the same algebraic structure underlies both the YB deformations and the conformal twists, regardless of the nature of the $r$-matrix, but only for unimodular $r$-matrices can we confidently invoke AdS/CFT. It is possible that a similar duality exists for solutions to generalized supergravity, as suggested originally in \cite{vanTongeren:2015uha}, but we will be conservative and not push this claim.

For the first one of the $r$-matrices~\eqref{r1,2},
\be
r_1 = \frac{1}{2}D \wedge K_1,
\ee
we note that it is a Jordanian $r$-matrix, $[D, K_1] = - K_1$, and the corresponding twist is a Drinfeld twist. However, it is clear that since the $r$-matrix is non-Abelian, in contrast to Abelian $r$-matrices e. g. \cite{Chaichian:2004za, Lukierski:2005fc}, there is not a simple expression for the corresponding twist element as the exponential of the $r$-matrix. Regardless, to leading order the twist element is
%we introduce the twist element $\hat{\mathcal{F}}$ as a differential operator, which acts on the commutative algebra $\mathcal{A}$ of functions, $f(x), g(x)$, in Minkowski space, following the standard procedure
%\footnote{Given a Lie algebra $\mathfrak{g}$ and its universal enveloping algebra ${\cal U}(\mathfrak{g})$, the Drinfeld twist ${\cal F}$ is defined as an invertible element of ${\cal U}(\mathfrak{g})\times {\cal U}(\mathfrak{g})$ that satisfies the cocycle condition $({\cal F}\otimes \boldsymbol{1})(\Delta\otimes \mathrm{id}){\cal F} = (\boldsymbol{1}\otimes {\cal F})(\mathrm{id}\otimes \Delta){\cal F}$, which is translated into the associativity of the star-product. Therefore, twists defined by ${\cal F}={\rm e}^{c_0 r_{na}}$, where $r_{na}$ is a nonabelian $r$-matrix, for instance (\ref{twist}), are not Drinfeld twists in the usual sense, because the cocycle condition is not satisfied in general. Moreover, the associativity of the star-product, as in (\ref{star}), must be defined as a supplementary condition. }
%\cite{Chaichian:2004za, Lukierski:2005fc}:  \red{correct this!}
\be
\hat{\mathcal{F}} \simeq  1  -i \eta \hat{D} \wedge \hat{K}_1 + \dots.
\ee
and this is enough to confirm that we recover the geometric result at the same order. 
The star product then takes the form,
\bea
\label{star}
f(x) \star g(x) &=& m \circ \hat{\mathcal{F}} (f(x) \otimes g(x)) = m \circ (1  -i \eta \hat{D} \wedge \hat{K}_1) (f(x) \otimes g(x)), \nn
&=& m \circ (1 + i \eta  (x^\rho \partial_\rho )  \wedge (2 x_1 \,x^\lambda \partial_\lambda - x^\lambda x_\lambda \, \partial_1) )( f(x) \otimes g(x)),
\eea
where $m$ denotes the operation of commutative multiplication, $m (f(x) \otimes g(x) ): = f(x) g(x)$. Note, that there is no $z$-dependence and the operators are essentially the AdS$_5$ Killing vectors evaluated at $z =0$. Taking $f(x) = x^{\mu}, g(x) = x^{\nu}$, $\mu, \nu = +, -, 1, 2$, while expanding to first order, one finds,
\bea
x^{\mu} \star x^{\nu} &=& x^{\mu} x^{\nu} -  \frac{i}{2} \eta (x^\lambda x_\lambda ) (x^\mu \delta_1^\nu - x^\nu \delta_1^\mu), \nn
x^{\nu} \star x^{\mu} &=& x^{\nu} x^{\mu} - \frac{i}{2} \eta (x^\lambda x_\lambda) (x^\nu \delta_1^\mu - x^\mu \delta_1^\nu).
\eea
Therefore, the Moyal bracket is
\be
[ x^{\mu}, x^{\nu} ]_{\star} = x^{\mu} \star x^{\nu} - x^{\nu} \star x^{\mu} =  - i \eta (x^\lambda x_\lambda)(x^\mu \delta_1^\nu - x^\nu \delta_1^\mu).
\ee
It is easy to read off the non-zero components of $\Theta^{\mu \nu}$:
\be
\Theta^{\mu 1} = -\eta x^\mu (x^\lambda x_\lambda).
\ee
This precisely agrees with the expression derived from the geometry (\ref{Theta-r1}), further evaluated at $z=0$.

Next, let us consider the second $r$-matrix,
\be
r_2 = \frac{1}{2} (P_0 - P_3) \wedge (D + M_{03}),
\ee
which is Abelian and can be exponentiated to get the twist element. 
It is convenient to introduce null coordinates, $x^{\pm} = x^0 \pm x^3$, so that the 4D Minkowski metric is
\be
\dd s^2 = - \dd x^+ \dd x^- + (\dd x^1)^2 + ( \dd x^2)^2,
\ee
where $\eta_{+ - } = - \frac{1}{2}, \eta^{+-} = -2$.
In these coordinates, the generators correspond to differential operators:
\be
\hat{P}_0 - \hat{P}_3 = - 2 \partial_{-}, \quad \hat{D} + \hat{M}_{03} = - 2 x^+ \partial_{+} - x^1 \partial_1 - x^2 \partial_2,
\ee
where again we have restricted the Killing vector fields to the boundary at $z=0$.

Using the Drinfeld twist
\be
\hat{\mathcal{F}} = e^{-2 i \eta \hat{r}_2} =  e^{-i \eta (\hat{P}_0 - \hat{P}_3) \wedge (\hat{D} + \hat{M}_{03})},
\ee
the star product can be written as
\bea
f(x) \star g(x) &=& m \circ \hat{\mathcal{F}} (f(x) \otimes g(x)) = m \circ e^{-i \eta (\hat{P}_0 - \hat{P}_3) \wedge (\hat{D} + \hat{M}_{03})} (f(x) \otimes g(x)), \nn
&=& m \circ e^{-i \eta  \partial_{-}  \wedge (2 x^+ \partial_{+} + x^1 \partial_1 + x^2 \partial_2)}( f(x) \otimes g(x)).
\eea
Specifying again $f(x) = x^{\mu}, g(x) = x^{\nu}$, one finds,
\bea
x^{\mu} \star x^{\nu} &=& x^{\mu} x^{\nu} -  \frac{i}{2} \eta (x^+ \eta^{\mu +} \eta^{\nu -}  - x^1 \eta^{\mu +} \eta^{\nu 1} - x^2 \eta^{\mu +} \eta^{\nu 1} - \mu \leftrightarrow \nu), \nn
x^{\nu} \star x^{\mu} &=& x^{\nu} x^{\mu} - \frac{i}{2} \eta (x^+ \eta^{\nu +} \eta^{\mu -}  - x^1 \eta^{\nu +} \eta^{\mu 1} - x^2 \eta^{\nu +} \eta^{\mu 1} - \nu \leftrightarrow \mu).
\eea
Note that in this case the $\eta$-expansion terminates at first order. Thus, the Moyal bracket is
\be
[ x^{\mu}, x^{\nu} ]_{\star} = x^{\mu} \star x^{\nu} - x^{\nu} \star x^{\mu} =  - i \eta (x^+ \eta^{\mu +} \eta^{\nu -}  - x^1 \eta^{\mu +} \eta^{\nu 1} - x^2 \eta^{\mu +} \eta^{\nu 1} - \mu \leftrightarrow \nu).
\ee
It is easy to read off the non-zero components of $\Theta^{\mu \nu}$:
\be
\Theta^{-+} = - 4  \eta x^+, \quad \Theta^{-1} = - 2  \eta x^1 , \quad \Theta^{-2} = - 2  \eta x^2.
\ee
This precisely agrees with the expression derived from the geometry (\ref{Theta-r2}), further evaluated at $z=0$.

 \section{$\Lambda$-symmetry and divergence of $\Th$}
 \label{sec:unimodular}
 In this section we motivate (\ref{unimodular}) from a symmetry principle. We will confine our attention to Yang-Baxter deformations, but recent applications of (\ref{unimodular}) to non-Abelian T-duality \cite{Fernandez-Melgarejo:2017oyu}, and in particular Ricci-flat Bianchi cosmologies \cite{Gasperini:1993nz}, where there is no D-brane description, suggest that a more general explanation should exist. In short, (\ref{unimodular}) is applicable even in the absence of a D-brane description, but here we present an explanation based on D-branes, which ultimately may also be inadequate. With this disclaimer, we proceed.

To begin, we review $\Lambda$-symmetry. We recall the string theory $\sigma$-model action in the Polyakov form,
\be
S_{\textrm{closed}}=\frac{1}{4\pi\alpha'}\int_\Sigma \dd^2 x\, \left (h^{ab}g_{MN}(X)\partial_a X^M \partial_b X^N+\epsilon^{ab}B_{MN}(X)\partial_a X^M \partial_b X^N\right).
\ee
We remark that under the $\Lambda$-transformation,
\be\label{Lambda-B}
B \to B + \dd \Lambda,
%B_{MN}\to B_{MN}+ (\dd \Lambda)_{MN},
\ee
where $\Lambda$ is a generic one-form,  the action transforms by a total derivative term,
$$
\delta_\Lambda S_{\textrm{closed}}=\frac{1}{2\pi\alpha'}\int_\Sigma \dd^2 x\, \epsilon^{ab}\partial_a\left(\Lambda_{M}(X)\,\partial_b X^M\right).
$$
As a result of this observation, we conclude that \eqref{Lambda-B} is a gauge symmetry of closed string theory. The $\Lambda$-symmetry is explicitly seen in the low-energy effective theory of closed strings in the fact that the $B$-field appears in the supergravity action through the field strength $ H = \dd B$.

When we are dealing with open strings, e.g. in the presence of D-branes, this $\Lambda$-symmetry is modified: the $\Lambda$-variation of the action is not zero any more. However, in these cases the worldsheet action has a boundary term \cite{Leigh:1989jq},
\be
S_{\textrm{boundary}}=\frac{1}{2\pi\alpha'}\int_{\partial\Sigma} dx^a \, A_M(X)\,\partial_a X^M,
\ee
where $A(X)$ is the gauge field along the D-brane.  Now, if together with \eqref{Lambda-B} we also transform $A$ as,
\be\label{Lambda-A}
A \to A-\Lambda,
\ee
the worldsheet theory retains $\Lambda$-invariance. This symmetry is also reflected in the low energy theory of D-branes, the DBI action \cite{Leigh:1989jq}, where the $B$-field appears only through $B+\dd A$ combination \cite{Witten:1995im, Leigh:1989jq,SheikhJabbari:1999ba}.

\subsection{$\Lambda$-symmetry and NC description}
We will now see how the condition~\eqref{unimodular} emerges as the equation of motion of the auxiliary gauge field $A$. One has to fix the gauge $\dd A=0$ after varying the action, in order to facilitate the replacement $(g+B)^{-1}{}^{[MN]} = \Th^{MN}$. Then, recalling the identity \cite{Seiberg:1999vs},
\be\label{G-g}
G_s^{-1} \sqrt{\det{G}} = g_s^{-1} {\rm e}^{-\Phi} \sqrt{\det{(g+B)}},
\ee
one has
\be\label{s1}\bn
\left. \left( \delta_{A} \int g_{s}^{-1} \, {\rm e}^{-\Phi} \sqrt{\det{(g + B + \dd A)}} \right)\right|_{\mathrm{d}A = 0} &= \frac{G_s^{-1}}{2} \int  \sqrt{\det G}\, (g+B)^{-1}{}^{MN} \partial_{[N} \delta A_{M]} \\
&= G_{s}^{-1} \int  \sqrt{ \det G}\, \nabla_M \Th^{MN} \,\delta A_{N}.
\en\ee
Note that $\nabla_M$ is the covariant derivative with respect to the Levi-Civita connection of the open string metric. In other words, the divergence of an antisymmetric tensor $\Theta^{MN}$ is computed using the open string metric,
\be
\nabla_M \Th^{MN} \equiv \frac{1}{\sqrt{ \det G}}\, \p_M \left( \sqrt{ \det G}\, \Th^{MN} \right).
\ee
The above then yields\footnote{A closely related argument may be found in  \cite{Szabo:2006wx}.}
\be\label{Theta-div-free}
\nabla_M\Th^{MN}=0.
\ee

%As discussed in the open string description, we may remove (a part of) the $B$-field and replace it with NC structure $\Theta$. 
A complementary viewpoint is to note that the open string parameters $G_{MN}$, $\Theta^{MN}$, $G_s$ are defined for a fixed $\Lambda$-gauge of the $B$-field. In other words, the NC description is in general $\Lambda$-gauge dependent \cite{Seiberg:1999vs,SheikhJabbari:1999ba}. There are of course specific actions that are invariant under (infinitesimal) $\Lambda$-transformations. Crucial observation is that deriving the equation of motion for $A$ in the gauge $dA=0$~\eqref{s1} is formally equivalent to varying the gauge fixed action with respect to $\Lambda$. To check which DBI actions are invariant, we impose that the LHS of~\eqref{G-g} be $\Lambda$-invariant: 
\be
\delta_{\Lambda} \int G_{s}^{-1} \, \sqrt{\det G} = \delta_{\Lambda} \int g_{s}^{-1} \, {\rm e}^{-\Phi}\sqrt{\det{(g+B)}} = G_{s}^{-1} \int  \sqrt{ \det G}\, \nabla_M \Th^{MN} \,\Lambda_{N} =0,
\ee
which yields the same constraint on $\Theta$.
%where we have set to zero the field strength of commutative and NC gauge fields. Requiring the LHS to be $\Lambda$-invariant implies that
%\be\label{s1}\bn
%\delta_{\Lambda}\left(\int g_{s}^{-1} \, {\rm e}^{-\Phi}\sqrt{\det{(g+B)}}\right)& =\\
%&= \frac12 \int  e^{-\f}\sqrt{\det(g+b)}\, \d_\LL \mathrm{Tr} \log (g+b) \\
%&= \frac12 \int  G_s^{-1} \sqrt{\det G}\, (g+B)^{-1}{}^{MN} \d_\LL B_{MN}
%\\&= \frac12\int  G_s^{-1} \sqrt{\det G}\, \Th^{\m\n} (\p_\n \LL_\m - \p_\m \LL_\n)\\
% &= \int  \p_\m \left( G_s^{-1} \sqrt{\det G}\, \Th^{\m\n} \right) \LL_\n\\
% &= G_{s}^{-1} \int  \sqrt{ \det G}\, \nabla_M \Th^{MN} \,\LL_{N} =0,
%\en\ee

We note that the full action governing the low-energy dynamics of the theory in the NC description is $S^{\textrm{NC}}_{\textrm{DBI}}+S_{\textrm{Sugra}}$. Since $S_{\textrm{Sugra}}$ is invariant under $\Lambda$-symmetry on its own, \eqref{Theta-div-free} is the condition for $\Lambda$-invariance of the full theory in the NC description. %Note also that, recalling \eqref{Lambda-A}, \eqref{Theta-div-free} is nothing but the equation of motion for the $A_M$ field.

The astute reader will note that none of the above discussion is tailored to a particular D-brane of given dimensionality and it is a generic observation that the LHS of (\ref{unimodular}) follows from varying a DBI action.  There is however an important distinction: $g, B$ are now closed string fields with dependence on holographic direction $z$, essentially the geometric data provided by YB deformations of AdS$_5$ geometries. Our analysis above works for any Dp-brane ($p\geq 3$) whose worldvolume includes the four directions on AdS$_5$ at a given $z$ while wrapping on some cycles on the interal deformed S$^5$ part.  For definiteness and simplicity, as we will do in the next section one may, however, focus on the case of a D3-brane at a fixed radial coordinate $z$, as is done in \cite{I-Sugra}. In the next section, we show that the RHS of (\ref{unimodular}) can be motivated from a Wess-Zumino term.

\subsection{Generalized supergravity and $\Lambda$-symmetry}

As discussed earlier, there are non-unimodular YB deformations, which correspond to background geometries that are not solutions to supergravity, but instead satisfy generalized supergravity equations of motion \cite{Arutyunov:2015mqj,Wulff:2016tju}. The generalized supergravity equations of motion (see appendix \ref{sec:gen_IIB}), besides the usual supergravity fields, involve an isometry (Killing) vector field $I^M$ \cite{Arutyunov:2015mqj,Wulff:2016tju}. Here, we revisit the $\Lambda$-invariance for generalized supergravity.

Going through the set of generalized supergravity equations, one can check that these equations are invariant under the $\Lambda$-transformation \eqref{Lambda-B} (see appendix \ref{sec:gen_IIB-inv}). Therefore, $\Lambda$-invariance of the generalized supergravity is guaranteed. In fact one can show that the modifications of the supergravity equations of motion from the closed string worldsheet theory viewpoint (i.e. cancellation of the 2d conformal anomaly and the $\beta$-function equations\footnote{The generalized supergravities are realized as special sections of Double Field Theory (DFT) \cite{Sakatani:2016fvh, Baguet:2016prz}. For argument on the Weyl invariance based on the DFT viewpoint, see \cite{Sakamoto:2017wor}.}) arise from requiring the worldsheet conformal invariance in presence of a given isometry vector field $I^M$. To see the latter one can choose a target space normal coordinate basis in which one of the target space directions is along $I^M$ and repeat the standard $\beta$-function analysis in this frame, while requiring vanishing of the Lie derivative of any supergravity field $X$, ${\cal L}_I X=0$.

The next step is to consider $\Lambda$-invariance of the DBI action in the presence of the modification $I^M$ introduced by generalized gravity. Here we focus on D3-branes that source the AdS$_5\times$S$^5$ geometry. More concretely, we recall that to consider D3-branes in the presence of background fields, we should consider an action comprising the DBI part and Chern-Simons, or Wess-Zumino, terms:
\be
S_{\textrm{DBI} + \textrm{WZ}} = \int_{\Sigma_4} \left( \frac{e^{- \Phi}}{g} \sqrt{ \textrm{det} (g + \mathcal{F} )} + \left(C_4 +  C_2 \mathcal{F} + \frac{1}{2} C_0  \mathcal{F}^2 \right) \right),
\ee
where we have defined $\mathcal{F} = F + B$, with $F$ being the field strength of the gauge field living on the brane, the D3-brane worldvolume $\Sigma_4$ and RR potentials $C_n$. For conciseness, we have omitted wedge products in the WZ term. For D3-branes embedded along the flat directions of the AdS$_5$ Poincar\'e patch metric, one finds a ``no force" condition, which is a hallmark of supersymmetric configurations.

Given the above action, there is a related action one can consider,
\be
S_{\textrm{AdS}_5} = \int_{\textrm{AdS}_5} \left( \frac{4 e^{- \Phi}}{g} \sqrt{ \textrm{det} (g + B)} - \left(F_5 + B \wedge F_3 + \frac{1}{2} B^2 F_1 \right) \right),
\ee
where $F_{n}$ are the usual RR field strengths and we have set the gauge fields to zero. This action is essentially the exterior derivative of the original D3-brane action and it is related to the previous action via Stokes theorem (see for example \cite{Schwarz:2013wra}). The factor of 4 and the minus sign appear to ensure the ``no force" condition for the AdS$_5\times$S$^5$ geometry supported by five-form flux. As explained in greater detail in \cite{I-Sugra}, varying this action with respect to $\Lambda$, while making use of the identities
\be
\label{identity}
\dd (F_3 + B \wedge F_1)  = i_{I} (F_5 + B \wedge F_3 + \frac{1}{2} B \wedge B \wedge F_1) = 4 i_{I} \vol (\textrm{AdS}_5),
\ee
and demanding that it is $\Lambda$-invariant, $\delta_{\Lambda} S_{\textrm{AdS}_5} = 0$, leads to the required equation on the nose,
\be
\label{Theta-I}
\nabla_{M} \Theta^{MN} = I^{N}.
\ee
While admittedly, our treatment here is a little quick, it should be noted that the first equality in (\ref{identity}) follows directly from the equations of motion of generalized supergravity, while the second equality can be understood as the well-known invariance of Page forms under TsT transformations. The reader will note that this is a non-trivial result where the variation of the Wess-Zumino term vanishes for pure supergravity backgrounds and results in an $I$ factor for generalized supergravity. We refer the reader to \cite{I-Sugra} for further details.

In summary, the take-home message is that (\ref{Theta-I}) is a consistency condition that arises from demanding that the D3-brane action, or more accurately its AdS$_5$ bulk counterpart, is invariant under $\Lambda$-symmetry when coupled to the background fluxes of (generalized) supergravity.

\section{Death by example}
\label{sec:examples}
In this section, we flesh out our prescription by discussing various examples. Where the solutions have already appeared in the literature, we omit the RR sector. In each case, from a knowledge of the NS sector of the YB deformation, we identify the NC structure from the geometry and show it satisfies (\ref{Theta-r}).  In cases where the YB deformed geometry is a supergravity solution, we confirm the NC structure is divergence-free, whereas if the resulting geometry is a solution to generalized IIB supergravity, we show that the divergence satisfies (\ref{unimodular}).

\subsection{A new unimodular example}
Before proceeding to YB deformations based on non-unimodular $r$-matrices, we consider deformations that are supergravity solutions. Having discussed the Maldacena-Russo geometry, it will be instructive for later purposes to consider the related Abelian twist
\begin{equation}
\label{eq:KK-r-matrix}
r=\frac{1}{2}K_1\wedge K_2\,.
\end{equation}
This $r$-matrix also satisfies the unimodularity condition (\ref{unimodular_r}), so one expects the output to be a supergravity solution. As we discuss later in section \ref{sec:auto}, one interesting feature of this $r$-matrix is that it is related via an outer automorphism of the conformal algebra to (\ref{PP}).  Owing to its length, we omit the corresponding expression in terms of differential operators, which may be easily worked out from (\ref{diff_operators}).

%Using the prescription given in section, augmented with  performing the supercoset construction~\cite{Kyono:2016jqy},
%the associated deformed background is obtained as
The YB deformation corresponding to the $r$-matrix may be written as \footnote{We have introduced $x_\m x^\m \equiv -x_0^2+x_1^2+x_2^2+x_3^2$ and $x_\m\bd x^\m\equiv-x_0 \bd x_0+x_1\bd x_1+x_2\bd x_2+x_3\bd x_3$.},
\begin{eqnarray}
\dd s^2&=&\frac{-\bd x_0^2+\bd x_3^2+\bd z^2}{z^2}
+\frac{z^2\left[\,\bd x_1^2+\bd x_2^2\,\right]}{z^4+\h^2(x_\m x^\m+z^2)^4}\no \\
&&+\frac{4\h^2(x_\m x^\m+z^2)^2}{z^2[\,z^4+\h^2(x_\m x^\m+z^2)^4\,]}
\biggl[-(x_1^2+x_2^2)\bigl(x_\m\bd x^\m+z\bd z\bigr)^2\no \\
&&+(x_\m x^\m+z^2)(x_1\bd x_1+x_2\bd x_2)\bigl(x_\m\bd x^\m+z\bd z\bigr)
\biggr]+\dd s^2(S^5)\,,\no\\
B &=&\frac{\h (x_\m x^\m+z^2)}{z^4+\h^2(x_\m x^\m+z^2)^4}
\biggl[2(x_1\bd x_2-x_2\bd x_1)\wedge (-x_0 \bd x_0 +x_3\bd x_3+z\bd z)\no \\
&&+\left(-x_0^2-x_1^2-x_2^2+x_3^2+z^2\right)\bd x_1\wedge \bd x_2
\biggr] \,, \quad
\Phi=\frac{1}{2}\log
\left[\frac{z^4}{z^4+\h^2(x_\m x^\m+z^2)^4}\right] \,,\nn
F_3 &=&\frac{8\h(x_\m x^\m+z^2)}{z^5}
\Bigl[\,\bd x_0 \wedge (x_1\bd x_1+x_2 \bd x_2)\wedge (-z\bd x_3+x_3\bd z)\no \\
&&+x_0 (x_1\,\bd x_1+x_2\,\bd x_2)\wedge \bd x_3\wedge \bd z +\frac{-x_0^2-x_1^2-x_2^2+x_3^2+z^2}{2}\,\bd x_0 \wedge \bd x_3\wedge \bd z\Bigr] \,,\no\\
F_5&=&4\left[\frac{z^4}{z^4+\h^2(x_\m x^\m+z^2)^4}\omega_{\textrm{AdS}_5}+\omega_{{\rm
		S}^5}\right]\,,
\label{KKbg}
\end{eqnarray}
where $\omega_{\textrm{AdS}_5}$ denotes the undeformed AdS$_5$ volume form. It is easy to confirm that the open string metric is undeformed with constant coupling, but the non-trivial information about the deformation resides in the NC structure,
\be\bn
\label{Theta-KK}
\Th^{0 1}  &=-2x_0 x_2 (x_\m x^\m + z^2) \h \,, \quad \Th^{0 2} = 2x_0 x_1 (x_\m x^\m + z^2) \h \,, \\
\Th^{1 2 } &=  (x_0^2 +x_1^2 + x_2^2 - x_3^2 -z^2)(x_\m x^\m + z^2) \h\,, \\
\Th^{1 3} &= 2 x_2 x_3 (x_\m x^\m + z^2)\h\,, \quad \Th^{1 z} = 2 x_2 z (x_\m x^\m + z^2) \h \,, \\
\Th^{2 3} &= -2 x_1 x_3 (x_\m x^\m + z^2) \h\,, \quad \Th^{2 z} = - 2 x_1 z(x_\m x^\m + z^2)\h \,.
\en\ee
It is lengthy, but nevertheless straightforward, to show that divergence vanishes in line with (\ref{unimodular}). Before leaving this example, we stress that this provides a non-trivial prediction for the corresponding Abelian twist of the conformal algebra. Once evaluated on the boundary, $z=0$, we expect the result to agree with the leading order NC deformation arising from a Drinfeld twist of the conformal algebra, thus extending earlier results on Poincar\'e twists \cite{Chaichian:2004za, Chaichian:2004yh, Lukierski:2005fc}. Since the above expression (\ref{Theta-KK}) is quartic, this extends the known $x$-dependence of the NC structure beyond quadratic level.

\subsection{Non-unimodular examples}
From this point onwards, we will focus on non-unimodular $r$-matrices solutions to the homogeneous cYBE, which will result in solutions to generalized supergravity. For each of these examples, the RHS of (\ref{unimodular}) is non-trivial, and we will check case by case that the equation is satisfied.

Let us first consider the non-Abelian classical $r$-matrix
\cite{vanTongeren:2015uha,Orlando:2016qqu}:
\be
\label{eq:space-r-matrix}
r=\frac{1}{2}P_1\wedge D\,.
\ee
Since $[D ,P_1] = P_1$, it is clear that this $r$-matrix can not satisfy the unimodularity condition \cite{Borsato:2016ose}. Therefore, the YB deformation is not a solution to the usual type IIB supergravity
but its generalization \cite{Arutyunov:2015mqj,Wulff:2016tju}.

The associated YB deformed geometry is given by
\begin{equation}
  \begin{aligned}
    \bd{s}^2 &= \frac{z^2[\bd{t}^2+\bd x_1^2+\bd{z}^2]+\eta^2(\bd{t}-t
      z^{-1}\bd{z})^2}{z^4+\eta^2(z^2+t^2)}
    +\frac{t^2(-\bd{\phi}^2+\cosh^2\phi \bd{\theta}^2)}{z^2}
    + \dd s^2(S^5), \\
    B &=-\eta\,\frac{t \bd{t} \wedge \bd{x^1} + z \bd{z}\wedge \bd{x^1}}{z^4+\eta^2(t^2+z^2)}\,, \quad
%    F_3 &= \frac{4\eta\, t^2\cosh\phi}{z^4}\left[\bd{t} \wedge
%      \dd{\theta} \wedge \dd{\phi}
%      -\frac{t}{z}\dd{\theta}\wedge \bd{\phi} \wedge \bd{z}  \right]\,, \\
%    F_5 &=
%    4\left[\frac{z^4}{z^4+\eta^2(t^2+z^2)}\omega_{AdS_5}+\omega_{{\rm
%          S}^5}\right]\,,\\
    \Phi = \frac{1}{2}\log
    \left[\frac{z^4}{z^4+\eta^2(t^2+z^2)}\right]\,,
  \end{aligned}
\label{space}
\end{equation}
where we have omitted the RR sector, which may be found in the literature \cite{vanTongeren:2015uha,Orlando:2016qqu}, and we have further defined,
\begin{align}
\label{eq:cartesian-coord-x0x2x3}
x_0=t \sinh\phi\,,  \qquad x_2 = t \cosh\phi\cos\theta\,,
\qquad x_3 =t \cosh\phi\sin\theta\,.
\end{align}
The geometry is a solution to generalized type IIB supergravity, where the vectors describing the modification take the form,
\begin{align}
  I =-\frac{\eta\, z^2}{z^4+\eta^2(t^2+z^2)} \dd{x_1}\,, \qquad
 Z =-\frac{2\eta^2 t}{z^4+\eta^2(t^2+z^2)}\left(\dd{t}-\frac{t}{z}\dd{z}\right) \,.
\end{align}
From the solution, it is clear that $\partial_{x_1}$ is a Killing direction. Indeed, raising the index on $I$,
 $I^{M}=g^{MN}I_N$, we note that modulo a factor and sign, we recover this Killing vector,
\be
 I =-\eta \partial_{x_1}.
\ee
Using earlier results, it is easy to read off the corresponding NC structure,
\be
\Theta^{t x_1}=\eta t\,,\qquad
\Theta^{z x_1}=\eta z
\label{PDNC}\,.
\ee
and confirm that it satisfies the equation (\ref{unimodular}), or more concretely that
\be
\nabla_{M}\Theta^{M x_1}=-\eta= I^{x_1}\,.
\ee
Note, in contrast to  \cite{vanTongeren:2015uha}, where the NC structures carry no $z$-dependence, the NC structure we have defined gives rise to holographic noncommutativity, namely the noncommutativity extends into the holographic direction $z$. The $z$-dependence cannot be gauged away and plays a key role in ensuring that the equation (\ref{unimodular}) holds.

Next, we consider a related example, which has not appeared in the literature to date,
\begin{equation}
\label{K1D}
r=\frac{1}{2}K_1\wedge D\,.
\end{equation}
Once again we see that the $r$-matrix is non-Abelian, since $[D, K_1] = - K_1$. As we will discuss later in section \ref{sec:auto}, since the $r$-matrix (\ref{K1D}) is related to (\ref{eq:space-r-matrix}) through an outer automorphism of the conformal algebra, the YB deformations are related through a coordinate transformation. For the moment, we record the solution and will return to the precise relation to the YB deformation (\ref{space}) later. Using similar notation to before, the YB deformed geometry may be expressed as,
\be\bn
    \bd{s}^2 &= \frac{z^2[\bd{t}^2+\bd{x_1}^2+\bd{z}^2]+\eta^2(x_1^2+t^2+z^2)^2(\bd{t}-t
      z^{-1}\bd{z})^2}{z^4+\eta^2(t^2+z^2)(x_1^2+t^2+z^2)^2}\\
    &+\frac{t^2(-\bd{\phi}^2+\cosh^2\phi\,\bd{\theta}^2)}{z^2}
    + \dd s^2(S^5)\,,\\
    B &=\frac{\eta (x_1^2+t^2+z^2)}{z^4+\eta^2(t^2+z^2)(x_1^2+t^2+z^2)^2} \bd x_1\wedge(t\bd t+z\bd z)\,, \\
     \Phi &= \frac{1}{2}\log
    \left[\frac{z^4}{z^4+\eta^2(t^2+z^2)(x_1^2+t^2+z^2)^2}\right]\,, \\
    F_3 &=\frac{4\eta\, t^2(x_1^2+t^2+z^2)\cosh\phi}{z^4}\left[\bd{t} \wedge
      \bd{\theta} \wedge \bd{\phi}
      -\frac{t}{z}\bd{\theta}\wedge \bd{\phi} \wedge \bd{z}  \right]\,, \\
    F_5 &=
    4\left[\frac{z^4}{z^4+\eta^2(t^2+z^2)(x_1^2+t^2+z^2)^2}\omega_{AdS_5}+\omega_{{\rm
          S}^5}\right]\,.
\label{Kspace}
\en\ee
The geometry is again a solution of the generalized supergravity equations with defining vectors
\be\bn
  I &=-\frac{\eta z^2\left[((x^1)^2-t^2-z^2)\bd x^1+2x^1(z\bd{z}+t\bd{t})\right]}{z^4+\eta^2(t^2+z^2)((x^1)^2+t^2+z^2)^2}\,, \\
 Z &=-\frac{2\eta^2t((x^1)^2+t^2+z^2)^2 }{z^4+\eta^2(t^2+z^2)((x^1)^2+t^2+z^2)^2}\left(\bd t-\frac{t}{z}\bd z\right) \,.
\en\ee
Raising the index on $I$, we encounter the special conformal generator,
\be
I = \eta \left[ (t^2 - x_1^2 + z^2) \partial_{x_1} - 2 x_1 ( t \partial_t + z \partial_{z} ) \right],
%I^{x^1}=\eta(t^2-(x^1)^2+z^2)\,,\quad
%I^{t}=-2\eta x^1 t\,,\quad
%I^{z}=-2\eta x^1 z
\label{DKI}
\ee
written in slightly unusual coordinates (\ref{eq:cartesian-coord-x0x2x3}). Using earlier results, one can determine the NC structure,
\be
\Theta^{t x_1}=\eta t (t^2+x_1^2+z^2)\,,\quad
\Theta^{z x_1}=\eta z (t^2+x_1^2+z^2)\,,
\ee
and calculate the divergence,
\be\bg
\nabla_{M}\Theta^{M x_1}=\eta(t^2-x_1^2+z^2)=I^{x_1}\,,\\
\nabla_{M}\Theta^{M t}=-2\eta x_1 t=I^{t}\,,\quad
\nabla_{M}\Theta^{M z}=-2\eta x_1 z=I^{z}
\,.
\eg\ee

Next, we consider the non-Abelian $r$-matrix
\begin{equation}
  r = \frac{1}{2\sqrt{2}} (P_0-P_3) \wedge (D-M_{03})\,,
  \label{r-4.1}
\end{equation}
which results in the following YB deformation:
\be\bn
  \label{KY-sol}
    \bd{s^2} &= \frac{-2\bd{x^+}\bd{x^-} + \bd{\rho^2} + \rho^2\bd{\theta^2} + \bd{z^2}}{z^2}
    -\eta^2\left[\frac{\rho^2}{z^6}+\frac{1}{z^4}\right](\bd{x^+})^2 + \dd s^2(S^5)\,, \\
    B &= -\eta\left[\frac{\rho \bd{x^+} \wedge \bd{\rho}}{z^4}+\frac{1}{z^3}\bd{x^+}\wedge \bd{z} \right]\,, \quad  \Phi = \Phi_0 \text{ (constant)}\,. \\
    F_3 &= -4\eta\left [\frac{\rho^2}{z^5} \dd{x^+} \wedge \bd{\theta} \wedge
      \bd{z}
      +\frac{\rho}{z^4} \bd{ x^+}\wedge \bd{\rho} \wedge \dd{\theta} \right]\,, \\
    F_5 &= 4 (\omega_{AdS_5}+\omega_{{\rm S}^5}) \,,
   \en\ee
The geometry is once again a solution of the generalized supergravity equations with vectors
\begin{align}
I = I_M \bd{x}^M = \frac{2\eta}{z^2} \bd{x^+}\,,  \qquad Z_M = 0 = B_{MN}I^N\,.
\end{align}
Raising the index on $I$, we recover the Killing vector
\be
I =-2\eta\, \partial_{x^-}.
\ee

The NC structure is easily determined
\be
\Theta^{zx^-}=\eta z\,,\qquad
\Theta^{\rho x^-}=\eta \rho\,.
\label{KYNC}
\ee
and (\ref{unimodular}) can be shown to hold,
\be
\nabla_{M}\Theta^{M x^-}=-2\eta=I^{x^-}\,.
\ee

Moving along, we consider a slightly more involved non-Abelian $r$-matrix
\begin{equation}
  r = \frac{1}{2\sqrt{2}}\Bigl[(M_{01}-M_{31})\wedge
  P_1+(M_{02}-M_{32})\wedge P_2+M_{03}\wedge(-P_0+P_3)\Bigr]\,.
\end{equation}
The associated YB deformed geometry may be expressed as,
\begin{equation}
  \begin{aligned}
    \dd{s}^2 ={}&\frac{1}{z^4-\eta^2(x^+)^2}\biggl[z^2(-2\dd{x^+}\dd{x^-}+\dd{z}^2)
    +2\eta^2z^{-2}x^+\rho \dd{x^+}\dd{\rho}-\eta^2 z^{-2}\rho^2(\dd{x^+})^2 \\
    &-\eta^2(x^+)^2z^{-2}\dd{z}^2\biggr]+\frac{\dd{\rho}^2+\rho^2\dd{\theta}^2}{z^2}
    +\dd s^2 (S^5)\,, \\
    B ={}&- \eta\,\frac{\dd{x^+} \wedge (\rho \dd{\rho}-x^+\dd{x^-})}{z^4-\eta^2(x^+)^2}\,, \quad   \Phi = \frac{1}{2}\log \left[\frac{z^4}{z^4-\eta^2(x^+)^2}\right]\,, \\
    F_3 ={}&- 4\eta\,\frac{\rho}{z^5}\left(\rho \dd{x^+}\wedge
      \dd{\theta}\wedge \dd{z}
      - x^+\dd{\rho}\wedge \dd{\theta} \wedge \dd{z}\right)\,, \\
    F_5 ={}&4\left[\frac{z^4}{z^4-\eta^2(x^+)^2}\omega_{AdS_5}+\omega_{{\rm
          S}^5}\right]\,,
  \end{aligned}
\end{equation}
The background is a solution of the generalized supergravity equations with defining vectors
\begin{align}
  I &=-\frac{3\eta\, z^2}{z^4-\eta^2 (x^+)^2}\dd{x^+}\,, &
                                                          Z &=-\frac{2\eta^2x^+}{z^4-\eta^2 (x^+)^2}\dd{x^+}-\frac{2\eta^2(x^+)^2}{z(z^4-\eta^2 (x^+)^2)}\dd{z} \,.
\end{align}

Raising the index on $I$, we find the simple Killing vector,
\be
 I =3\eta \partial_{x^-}.
\ee

The NC structure is given by
\be
\Theta^{x^+x^-}=\eta x^+\,,\qquad
\Theta^{\rho x^-}=\eta \rho\,,
\ee
which allows one to again confirm (\ref{unimodular}),
\be
\nabla_{M}\Theta^{M x^-}=3\eta= I^{x^-}\,.
\ee
%Finally, we note that the vector $I^M$ can also be expressed as in \eqref{I-K} with
%$$
%{\cal K}=({\rm constant})\times {x^+ \rho}^{-3/2}\propto \Theta^{x^+x^-}\Theta^{\rho x^-}.
%$$

As our final example, we study the  classical $r$-matrix,
\begin{equation}
  r = -D\wedge P_0-M_{0\mu} \wedge P^{\mu}-M_{12} \wedge P_2-M_{13} \wedge P_3\,. \label{HvT-r}
\end{equation}
which was originally studied in~\cite{Hoare:2016hwh} in relation to a scaling limit of the classical $r$-matrix of Drinfeld-Jimbo type.
The resulting YB deformed geometry is given by
\begin{equation}
  \label{HvT}
  \begin{aligned}
    \dd{s}^2 &= \frac{-\dd{x_0}^2+\dd{z}^2}{z^2-4\eta^2}
    +\frac{z^2\left[\dd{x_1}^2+\dd{\rho}^2\right]}{z^4+4\eta^2\rho^2}
    +\frac{\rho^2\dd{\theta}^2}{z^2}+ \dd{s}^2_{{\rm S}^5}\,,  \\
    B &= -\frac{2\eta }{z(z^2-4\eta^2)} \dd{x_0}\wedge \dd{z}
    -\frac{2\eta\, \rho}{z^4+4\eta^2 \rho^2} \dd{x_1}\wedge \dd{\rho}\,, \\
%    F_{1} &= \frac{16\eta^2\rho^2}{z^4}\dd{\theta}\,, \\
%    F_{3} &=- \frac{8\eta \rho^2}{z^3(z^2-4\eta^2)} \dd{x^0}\wedge
%    \dd{\theta} \wedge \dd{z}
%    -\frac{8\eta \rho}{z^4+4\eta^2\rho^2} \dd{x^1}\wedge \dd{\rho} \wedge \dd{\theta}\,,  \\
%    F_{5} &=
%    4\left[\frac{z^6}{(z^2-4\eta^2)(z^4+4\eta^2\rho^2)}\,\omega_{AdS_5}
%      +\omega_{{\rm S}^5}\right] \,, \\
    \Phi &=
    \frac{1}{2}\log\left[\frac{z^6}{(z^2-4\eta^2)(z^4+4\eta^2\rho^2)}\right]\,,
  \end{aligned}
\end{equation}
where we have omitted the RR sector. This geometry is a solution of the generalized supergravity equations with the following vectors :
\begin{align}
I &= \frac{8\eta \dd{x^0}}{z^2-4\eta^2}-\frac{4z^2\eta \dd{x^1}}{z^4+4\eta^2\rho^2}\,, &
Z &= \left[\frac{2(z^2+2\eta^2)}{z(z^2-4\eta^2)}-\frac{2z^3}{z^4+4\eta^2\rho^2}\right]\dd{z}
+\frac{4\eta^2\rho \dd{\rho}}{z^4+4\eta^2\rho^2}\,. \label{I-6}
\end{align}
Raising the index on $I$, we identify the constant Killing vector,
\be
I = - 4 \eta ( 2 \partial_{x_0} + \partial_{x_1} )\,.
%I^{x^0}=-8\eta\,,\qquad
%I^{x^1}=-4\eta
\ee

Once determined, the NC structure takes the form,
\be
\Theta^{zx^0}=2\eta z\,,\qquad
\Theta^{\rho x^1}=-2\eta \rho\,.
\label{HTNC}
\ee
and calculating the divergence, one may again confirm that (\ref{unimodular}) is satisfied,
\be
\nabla_{M}\Theta^{M x^0}=-8\eta=I^{x^0}\,,\qquad
\nabla_{M}\Theta^{M x^1}=-4\eta=I^{x^1}
\,.
\ee

\section{Automorphism, YB deformations \& self-T-duality}
\label{sec:auto}
As remarked in  \cite{Borsato:2016ose}, since inner automorphisms of the algebra correspond to field redefinitions in the string sigma model, they are expected to be coordinate transformations in the target space. In contrast, in this section, we consider everyone's favorite outer automorphism of the conformal algebra,
\be
\label{outer_auto}
D \leftrightarrow - D, \quad P_{\mu} \leftrightarrow K_{\mu}.
\ee
This automorphism of the conformal algebra may be implemented through the coordinate transformation, \footnote{This transformation appears not to be widely known, but see for example \cite{Beisert:2010kp}.}
\be
\label{outer_auto_coord}
x^{\mu} \rightarrow \frac{x^{\mu}}{(x_{\nu} x^{\nu} + z^2)}, \quad z \rightarrow \frac{z}{(x_{\nu} x^{\nu} + z^2)}.
\ee
In order to better understand this transformation, consider the action of special conformal transformations on AdS coordinates:
\bea
x^{\mu} \rightarrow {x'}^{\mu} = \frac{x^{\mu} + b^{\mu} ( x_{\nu} x^{\nu} + z^2)}{1 + 2 b \cdot x + b^2 (x_{\nu} x^{\nu} + z^2) }, \quad z \rightarrow z' = \frac{z}{1 + 2 b \cdot x + b^2 (x_{\nu} x^{\nu} + z^2)}.
\eea
From here we see that the action of the special conformal transformation is mapped to a shift using the transformation (\ref{outer_auto_coord}),
\be
\frac{{x'}^{\mu}}{{x'}_{\nu} {x'}^{\nu} + {z'}^2} = \frac{x^{\mu}}{x_{\nu} x^{\nu} + z^2} + b^{\mu}.
\ee
To see that (\ref{outer_auto_coord}) is an automorphism, it is easy to check that it leaves the AdS metric invariant. To further confirm this, one can use the representation of $P_{\mu}, K_{\mu}$ and $D$ as differential operators (see appendix~\ref{sec:conventions});  it is easy to see that~\eqref{outer_auto_coord} maps the differential operators to each other. We now consider two applications.

\subsection{YB deformations}
For our first application, we will show that this automorphism can be used to generate new YB deformations. For example, consider the Maldacena-Russo geometry (\ref{MR}), which corresponds to the Abelian $r$-matrix $r = \frac{1}{2} P_1 \wedge P_2$. Invoking the automorphism of the algebra, one can map this $r$-matrix to $ r = \frac{1}{2} K_1 \wedge K_2$. It is then expected that the resulting geometries, namely (\ref{MR}) and (\ref{KKbg}), are related through the coordinate transformation (\ref{outer_auto_coord}). It is straightforward to check this is indeed the case.

This transformation indeed percolates through the rest of our story. Consider the constant NC structure corresponding to the Maldacena-Russo geometry, $\Theta^{1 2} = - \eta$. On the contrary, the corresponding NC structure for the geometry (\ref{KKbg}) is (\ref{Theta-KK}).  Admittedly this expression is suitably involved, but recalling that it transforms as a tensor,
\be
\Th^{M N} = \frac{\partial x^{M} }{\partial \tilde{x}^{M} } \frac{\partial x^{N}}{\partial \tilde{x}^{N}}  \Th^{\tilde{M} \tilde{N}},
\ee
one can show that it also may be mapped back to the simple constant NC structure through (\ref{outer_auto_coord}). We expect similar observations to hold for YB deformations where the $r$-matrices are related through the outer automorphism (\ref{outer_auto}). Moreover, our divergence conditions on $\Theta^{MN}$ and the holographic noncommutativity feature are covariant equations and  hence remain valid under coordinate transformation (\ref{outer_auto_coord}).

\subsection{Self-T-duality}
It is known that the maximally symmetric spaces AdS$_p\times$S$^p$, $p= 2, 3, 5$ \cite{Berkovits:2008ic, Beisert:2008iq, Adam:2009kt,  Dekel:2011qw, OColgain:2012ca}, and geometries based on exceptional Lie supergroups AdS$_q\times$S$^q\times$S$^q$ \cite{Abbott:2015mla, Abbott:2015ava} $q= 2, 3$, may be mapped back to themselves under a combination of bosonic and fermionic T-dualities. In the process of performing this transformation, the non-Lagrangian ``dual superconformal symmetry" \cite{Drummond:2006rz} is mapped back to the original superconformal symmetry of the theory, and combined they give rise to Yangian symmetry \cite{Drummond:2009fd}, a recognised  structure for integrablility. As a result, it is expected that all self-T-dual geometries are classically integrable, but the precise relation is unknown. \footnote{It is worth noting that there are integrable geometries (AdS$_4\times \mathbb{CP}^3$) that are not self-T-dual \cite{Adam:2010hh, Bakhmatov:2010fp, Colgain:2016gdj}.}

In this subsection, we study self-T-duality of AdS$_p\times$S$^p$ geometries, but with a small twist. Instead of using the accustomed isometries, we will instead T-dualize with respect to isometries that are related through the automorphism (\ref{outer_auto}), but extended to the full superconformal algebra. To be concrete, we will work in the supergravity description (reviewed in appendix \ref{sec:review_ftduality}),
with the Killing spinors from Poincar\'e patch \cite{Lu:1998nu}
\be
\eta = z^{-\frac{1}{2}} \eta_{Q} +  \left[ z^{\frac{1}{2}} + z^{-\frac{1}{2}} x^{\mu} \Gamma_{\mu} \right] \eta_{S}, \quad \Gamma_{z} \eta_{Q}  = -\eta_{Q}, \quad \Gamma_{z}  \eta_S = \eta_S,
\ee
where we have used subscripts $Q$ and $S$ to distinguish Poincar\'e and superconformal Killing spinors, respectively. It should be noted that the coordinate dependence is very different. Instead of following the usual prescription, where $\eta_S =0$, here we will instead set $\eta_{Q} = 0$ when performing fermionic T-dualities.

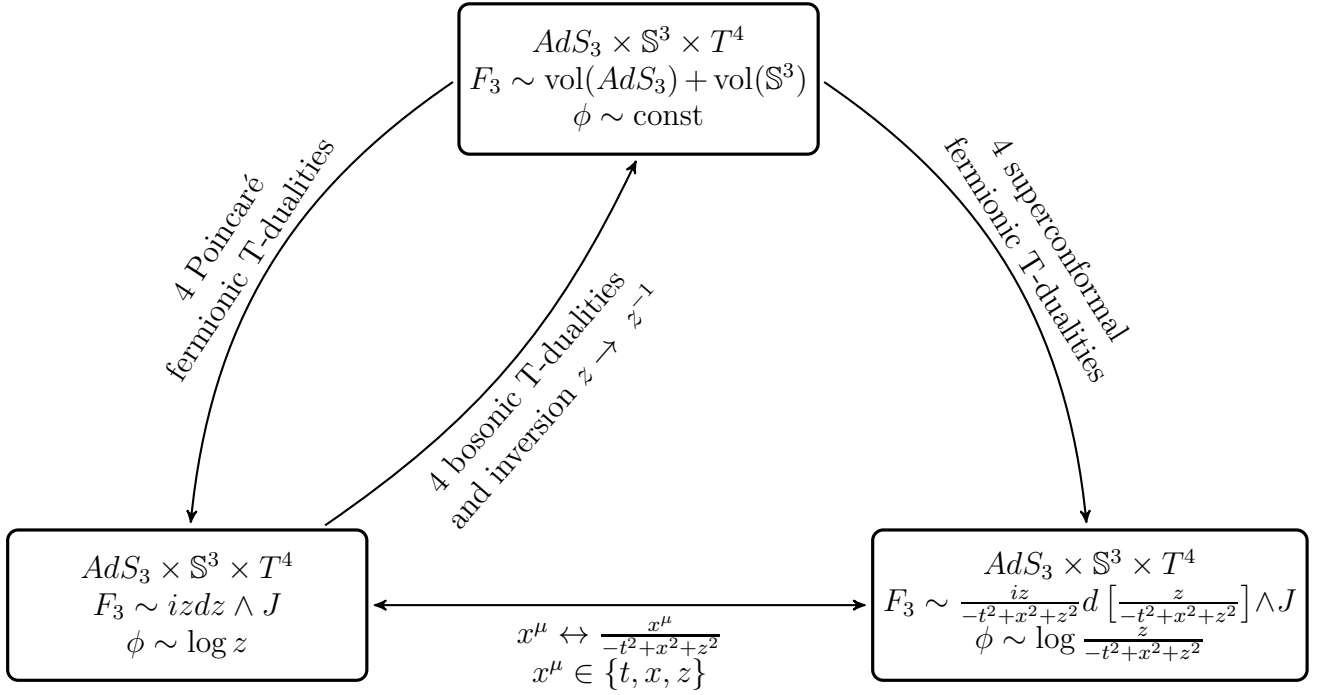
\begin{figure}
\centering
\begin{tikzpicture}

\node[format] at (6,0) (1) {\parbox{4.5cm}
							    {
								\centering $AdS_3 \times \mathbb{S}^3 \times T^4$\\
								$F_3 \sim \mathrm{vol}(AdS_3) + \mathrm{vol}(\mathbb{S}^3)$\\
								$\phi \sim \mathrm{const}$
								}
							 };

\node[format] at (0,-7)  (2) {\parbox{4.5cm}
								{
								\centering $AdS_3 \times \mathbb{S}^3 \times T^4$\\
								$F_3 \sim i z dz \wedge J$\\
								$\phi \sim \log z$
								}
							 }
	
	edge[arrow,<-,bend left=25] node[midway,rotate=57,above]
	{\parbox{4cm}{\centering 4 Poincar\'e \\ fermionic T-dualities}}
	(1.west)
	
	edge[arrow,->,bend right=15] node[midway,rotate=49,below]
	{\parbox{5cm}{\centering 4 bosonic T-dualities \\ and~inversion~$z\rightarrow~z^{-1}$}}
	(1.south);

\node[format] at (12,-7) (3) {\parbox{5.5cm}
								{
								\centering $AdS_3 \times \mathbb{S}^3 \times T^4$\\
								$F_3 \sim \frac{iz}{-t^2+x^2+z^2} d \left[\frac{z}{-t^2+x^2+z^2}\right] \wedge J$\\
								$\phi \sim \log \frac{z}{-t^2+x^2+z^2}$
								}
							 }
	
	edge[arrow,<-,bend right=25] node[midway,rotate=-58,above]
	{\parbox{4cm}{\centering 4 superconformal \\ fermionic T-dualities}}
	(1.east)
	
	edge[arrow,<->] node[midway,below]
	{\parbox{3cm}{\centering $x^\mu \leftrightarrow \frac{x^\mu}{-t^2 + x^2 + z^2}$\\ $x^\mu \in \{t,x,z\}$}}
	(2);

\end{tikzpicture}
\caption{Scheme of $AdS_3 \times S^3 \times T^4$ self-T-duality.} \label{fig}
\end{figure}

The idea is summarized in Figure~\ref{fig}. It is easiest to begin our study with the AdS$_3\times$S$^3\times$T$^4$ geometry, since the Killing spinors corresponding to commuting fermionic isometries have been previously identified in \cite{OColgain:2012ca}. We will see that one can use the same Killing spinors with $\eta_Q$ simply replaced by $\eta_S$. To get oriented, let us consider the solution, which we can explicitly write as
\bea
\label{geometry} \dd s^2 &=& \frac{- \dd x_0^2 + \dd x_1^2 + \dd z^2}{z^2} + \dd \theta^2 + \sin^2 \theta(\dd \phi^2 + \sin^2 \phi \dd \psi^2) + \dd s^2(T^4), \\
F_3 &=& \frac{2}{z^3} \dd x_0 \wedge \dd x_1 \wedge \dd z + 2 \sin^2 \theta \sin \phi \dd \theta \wedge \dd \phi \wedge \dd \psi,
\eea
where we are using nested coordinates for the S$^3$ metric and the Poincar\'e metric for AdS$_3$.

Following  \cite{OColgain:2012ca}, we introduce a basis for the Killing spinors so that the required projection condition $\Gamma^{6789} \e = - \e$ is imposed. We consider the same complex basis spinors
\bea
\xi_{a} = \left( \begin{array}{c} 1 \\ i \end{array} \right) \otimes \chi_{a} \otimes \left(  \begin{array}{c} 1 \\ i \end{array} \right), \quad \xi_{a+4} = \left( \begin{array}{c} 1 \\ -i \end{array} \right) \otimes \chi_{a} \otimes \left(  \begin{array}{c} 1 \\ -i \end{array} \right),
\eea
where $\chi_{a}$, $a = 1, \dots, 4$, denote four-component spinors of the form $\chi_{1} = (1, 0, 0, 0)^{T}$, $\chi_{2} = (0, 1, 0, 0)^{T}$, etc. Making use of the explicit real gamma matrices in appendix A of  \cite{OColgain:2012ca}, one can rewrite the Killing spinors in terms of 16 x 16 gamma matrices so that the Weyl spinors take the form:
\be\bn
\e &= \left[ z^{\frac{1}{2}} + z^{-\frac{1}{2}} ( -x_0 \, \gamma^{2} + x_1 \, \gamma_{1}^{~2} )\right] e^{\frac{\theta}{2} \gamma^{1 45}} e^{ \frac{\phi}{2} \gamma^{34}} e^{\frac{\psi}{2} \gamma^{45}} \eta_S, \\
\hat{\e} &= - \left[ z^{\frac{1}{2}} + z^{-\frac{1}{2}} ( - x_0 \, \gamma^{2} + x_1 \, \gamma_{1}^{~2} )\right] \gamma^{1} e^{\frac{\theta}{2} \gamma^{1 45}} e^{ \frac{\phi}{2} \gamma^{34}} e^{\frac{\psi}{2} \gamma^{45}} \eta_S.
\en\ee
It was found for Poincar\'e supersymmetries that the fermionic isometries built from the constant Killing spinors,
\be
\eta_1 = \xi_1 + \xi_6, \quad \eta_2 = \xi_2 - \xi_5, \quad \eta_3 = \xi_3 + \xi_8, \quad \eta_4 = \xi_4 - \xi_7,
\ee
commute, in the sense that the constraint (\ref{constraint}) is satisfied. Simply replacing the Poincar\'e supersymmetries with superconformal supersymmetries, we find that the same constraint holds. This allows us to identify four commmuting fermionic isometries corresponding to superconformal supersymmetries.

Making use of the same basis spinors, modulo an overall factor, we recover a very similar matrix to (2.23) of  \cite{OColgain:2012ca}:
\bea
C = \frac{16 (-x_0^2+x_1^2 + z^2)}{z} \left( \begin{array}{cccc} i c_{\theta} - s_{\theta} s_{\phi} s_{\psi} & - i  s_{\theta} c_{\phi} & -  s_{\theta}  s_{\phi} c_{\psi} & 0 \\
	- i s_{\theta} c_{\phi} & - i c_{\theta} - s_{\theta} s_{\phi} s_{\psi} & 0 & - s_{\theta} s_{\phi} c_{\psi} \\
	- s_{\theta} s_{\phi} c_{\psi} & 0 & i c_{\theta} + s_{\theta} s_{\phi} s_{\psi} & - i s_{\theta} c_{\phi} \\
	0 & - s_{\theta} s_{\phi} c_{\psi} & - i s_{\theta} c_{\phi} & - i c_{\theta} + s_{\theta} s_{\phi} s_{\psi}
\end{array} \right), \nonumber
\eea
where we have employed the obvious shorthand $c_{\theta} = \cos \theta, s_{\theta} = \sin \theta$, etc. From the determinant (\ref{fermion_dilaton}), one calculates the shift in the dilaton,
\be
\Phi = \frac{1}{2} \log \textrm{det} C = 2 \log \left[ \frac{(-x_0^2 + x_1^2 +z^2)}{z} \right],
\ee
where we have dropped a constant, which can be safely done by rescaling the constant Killing spinors. Next, from the shift in the bi-spinor (\ref{fermion_RR}), one can read off the fluxes that support the fermionic T-dual geometry,
\be
F_3 =   \frac{2 \, i \,z}{(-x_0^2 + x_1^2 + z^2)} \dd \left[ \frac{z}{(-x_0^2 + x_1^2 +z^2)}\right] \wedge J,
\ee
where $J = \dd x_6 \wedge \dd x_8 - \dd x_7 \wedge \dd x_9$ is the K\"ahler form on the T$^4$, and the geometry (\ref{geometry}) is unchanged.

The key observation at this point is modulo the denominator, $(-x_0^2 + x_1^2 +z^2)$, this is simply the result from the transformation with respect to Poincar\'e supersymmetries. In fact, we can recover the result simply by performing the transformation (\ref{outer_auto_coord}). As a result of this transformation, the solution can be brought to the form,
\be\bn
\dd s^2 &= \frac{(- \dd x_0^2 + \dd x_1^2 + \dd z^2)}{z^2} + \dd \theta^2 + \sin^2 \theta(\dd \phi^2 + \sin^2 \phi \dd \psi^2) + \dd s^2(T^4), \\
F_3 &= 2 i z \dd z \wedge J, \quad \Phi = - 2 \log z.
\en\ee
At this stage restoring the geometry reduces to the example discussed in  \cite{OColgain:2012ca}; further T-dualizing on $x_0, x_1, x_6, x_8$ in turn, while inverting $z \rightarrow z^{-1}$, one recovers the original geometry (\ref{geometry}). The AdS$_5\times$S$^5$ self-T-duality transformation is similar and we present some details in appendix \ref{sec:review_ftduality}.

\section{Discussion}\label{discussion-section}
In this work, we fleshed out some of the details of the earlier letter \cite{Araujo:2017jkb}. We will summarize the new results, thus extending the scope of the original letter. Building on the observation that for each homogeneous YB deformation of AdS$_5$ there exists an undeformed open string metric, in this paper we showed that for generic TsT transformations, provided the {NS-NS} two-form vanishes, the TsT-deformed geometry may be mapped to an open string metric, which is simply the original undeformed metric. Moreover, the corresponding string coupling is the original string coupling and the NC structure is a constant. In the concrete context of homogeneous YB deformations of AdS$_5$ we proved that the NC structure is related to the $r$-matrix through the expression (\ref{Theta-r}). Our prescription only captures the linear order dependence of the noncommutativity structure on the deformation parameter $\eta$, as is demonstrated by comparison to generic Drinfeld twists.

In this paper, we coined the term ``holographic noncommutativity". In contrast to earlier works \cite{vanTongeren:2015uha, vanTongeren:2016eeb}, our prescription for reading off the NC structure leads to NC structures that depend on the holographic direction $z$ and can also have $\Theta^{z\mu}$ components. This prescription is the natural way to identify all the NC structures corresponding to YB deformations of the full conformal algebra, since ultimately generators of the conformal algebra map to Killing vectors in the bulk AdS$_5$ spacetime, which inherently are $z$-dependent. For unimodular $r$-matrices, i. e. those leading to supergravity solutions, one novel feature of holographic noncommutativity is that the NC structure of the dual field theory is recovered on the boundary. To understand this better, we note that not all $\Theta^{MN}$ components are independent. There are constraints among them \eqref{Theta-div-free} or \eqref{Theta-I}. With these constraints the number of independent components in $\Theta^{MN}$ is exactly the same as $\Theta^{\mu\nu}$ which is defined on the boundary of AdS. In other words, one can solve \eqref{Theta-div-free} or \eqref{Theta-I} over AdS$_5$ with $\Theta^{\mu\nu}$ at the boundary $z=0$ being the initial conditions for these first order equations. A generic result of this analysis is that the $\Theta^{z\mu}$ components vanish at $z=0$, as confirmed by our long list of examples. A similar statement is also true for the Killing vector field $I^M$: $I^M$ is uniquely specified in the bulk by giving $I^\mu$ components at $z=0$.

A latter part of this work concerns the explanation of the unimodularity condition from the open string perspective. For unimodular $r$-matrices, we traced the divergence-free condition on the NC structure to the fact that the theory is $\Lambda$-symmetric. Moreover, for non-unimodular $r$-matrices, where we are forced to go beyond the supergravity description, we have identified a remarkable equation connecting closed and open string descriptions (\ref{unimodular}). In this work, we show that this condition is satisfied for a large class of YB deformations corresponding to non-unimodular $r$-matrices we examined and studied here, thus instilling confidence that it should be correct for all cases. We also interpreted the Killing vector $I$ mathematically as the modular vector field for the NC structure $\Theta$ (see appendix \ref{appen:cohomology}).

This paper also contains some novel results related to an outer automorphism of the conformal algebra, which corresponds to a coordinate transformation in AdS. Firstly, we showed that the coordinate transformation connects YB deformations, where the $r$-matrices may be mapped to each other under the action of the automorphism. This is as expected and demonstrates that the YB machinery is consistent. Secondly, we showed that this automorphism plays a natural r\^ole when one maps AdS$_p\times$S$^p$ geometries back to themselves under a combination of fermionic T-dualities with respect to superconformal supersymmetries and compensating bosonic T-dualities along special conformal isometries. While this self-T-duality transformation is expected, our work provides the first illustration of the mapping.

We end by briefly discussing the physical significance and implications of our results for the integrability of the dual NC sYM theories. Our musings echo similar comments made earlier in \cite{vanTongeren:2015uha}. Let us first consider the closed string theory on AdS$_5\times$S$^5$ background. This worldsheet theory describes propagation of non-interacting closed strings on the background. The perturbative string interactions may then be introduced via vertex operators. The integrability of the closed string $\sigma$-model on the AdS$_5\times$S$^5$ background then corresponds to the planar integrability of the dual sYM side. (Note that according to standard AdS/CFT dictionary, the non-planar $1/N$ effects in the sYM side corresponds to strings interactions on the background. Note also that the ``direct'' planar integrability checks in the sYM side have been carried out in a perturbation series in the 't Hooft coupling and then extended to all orders in perturbation theory using the Yangian symmetry. The $\sigma$-model integrability, however, corresponds to the large 't Hooft coupling limit. The two planar sYM and closed string worldsheet integrabilities,  while confirming each other, are in a sense complementary to one another.) On the other hand, it is now an established interesting result that closed string $\sigma$-model integrability remains for AdS$_5\times$S$^5$ deformations generated by conformal twists. Moreover, as we discussed in this work, field theories dual to these closed strings are NC deformations of sYM with the noncommutativity specified by the twist. One immediate conclusion one may infer is that the NC sYM we discussed here should also remain planar integrable. This result for a special class of Abelian Poincar\'e twists was already mentioned in \cite{Beisert:2005if}. Our analysis provides further supportive evidence for this result, as well as extending it to larger class of conformal twists. It is of course desirable to back up this important result by direct analysis in the NC sYM side. We hope to pursue this in future work.

\section*{Acknowledgements}

We are grateful to Henrique Burstyn, C.~Y.~Park, A.~Torrielli, B.~Vicedo and M.~Wolf for discussion. I.B. is partially supported by the Russian Government programme for competitive growth of Kazan Federal University. E \'O C. acknowledges the University of Surrey and Galileo Galilei Institute for Theoretical Physics, through the program ``New developments in AdS$_3$/CFT$_2$ holography", for hospitality during the writing-up stage. The work of J.S.\ is supported by the Japan Society for the Promotion of Science (JSPS).  The work of  M.M.Sh-J. is supported in part by junior research chair in black hole physcis of the Iranian NSF, the Iranian SarAmadan federation grant and the ICTP network project NT-04.  K.Y.\ acknowledges the Supporting Program for Interaction-based Initiative Team Studies (SPIRITS) from Kyoto University and a JSPS Grant-in-Aid for Scientific Research (C) No.\,15K05051. This work is also supported in part by the JSPS Japan-Russia Research Cooperative Program.

\appendix

\section{Conformal algebra}
\label{sec:conventions}
For completeness, we record the conformal algebra $\frak{so}(4, 2)$ employed in this work,
\bea
\left[ D, P_{\mu} \right] &=& P_{\mu}, ~~ \left[ D, K_{\mu} \right] = - K_{\mu}, \nn
\left[ P_{\mu}, K_{\nu} \right]  &=&  2 \left( \eta_{\mu \nu} D +  M_{\mu \nu} \right), \\
\left[ M_{\mu \nu}, P_{\rho} \right] &=& {-} 2 \eta_{\rho [ \mu} P_{\nu]}, ~~  \left[ M_{\mu \nu}, K_{\rho} \right] = {-} 2 \eta_{\rho [ \mu} K_{\nu]}, \nn
\left[ M_{\mu \nu}, M_{\rho \sigma} \right] &=& {-} \eta_{\mu \rho} M_{\nu \sigma} {+} \eta_{\nu \rho} M_{\mu \sigma} {+} \eta_{\mu \sigma} M_{\nu \rho}  {-} \eta_{\nu \sigma} M_{\mu \rho}. \nonumber
\eea
The algebra can be realized in terms of differential operators as \footnote{In this paper, we distinguish the differential operators from the matrix generators by attaching the ``hat''.}
\bea
\label{diff_operators}
\hat{P}_{\mu} &=& -\partial_{\mu}, \quad
\hat{K}_{\mu} =  -(x_{\nu} x^{\nu}  +z^2) \partial_{\mu}
+ 2 x_{\mu} (x^{\nu} \partial_{\nu} + z \partial_z),
\nn
\hat{D} &=& - x^{\mu} \partial_{\mu}  - z \partial_{z}, \quad \hat{M}_{\mu \nu} =  x_{\mu} \partial_{\nu} {-} x_{\nu} \partial_{\mu}.
\eea

Following  \cite{Kyono:2016jqy}, we define the following matrix representations for the generators. The generators of the conformal algebra may be expressed as
\bea
P_{\mu} = \frac{1}{2} (\gamma_{\mu} - \gamma_{\mu} \gamma_5), \quad M_{\mu \nu} = \frac{1}{4} [ \gamma_{\mu}, \gamma_{\nu} ], \quad D = \frac{1}{2} \gamma_5, \quad K_{\mu} = \frac{1}{2} ( \gamma_{\mu} + \gamma_{\mu} \gamma_5),
\eea
where $\mu = 0, 1, 2, 3$. It is convenient to adopt the following realization of the gamma matrices,
\bea
\gamma_0 &=& \left( \begin{array}{cccc} 0 & 0 & 1 & 0 \\ 0 & 0 & 0 & -1 \\ -1 & 0 & 0 & 0 \\ 0 & 1 & 0 & 0 \end{array} \right), \quad \gamma_1 = \left( \begin{array}{cccc} 0 & 0 & 0 & -1 \\ 0 & 0 & 1 &0 \\ 0 & 1 & 0 & 0 \\ -1 & 0 & 0 & 0 \end{array} \right), \quad \gamma_2 = \left( \begin{array}{cccc} 0 & 0 & 0 & i \\ 0 & 0 & i & 0 \\ 0 & -i & 0 & 0 \\ -i & 0 & 0 & 0 \end{array} \right), \nn
\gamma_3 &=& \left( \begin{array}{cccc} 0 & 0 & 1 & 0 \\ 0 & 0 & 0 & 1 \\ 1 & 0 & 0 & 0 \\ 0 & 1 & 0 & 0 \end{array} \right), \quad \gamma_5 = \left( \begin{array}{cccc} 1 & 0 & 0 & 0 \\ 0 & 1 & 0 & 0 \\ 0 & 0 & -1 & 0 \\ 0 & 0 & 0 & -1 \end{array} \right).
\eea

${\bf P}_{m}$, which feature in the definition of the  YB $\sigma$-model in section \ref{sec:YBreview}, may be defined in terms of the gamma matrices as,
\be
{\bf P}_{m} =  - \frac{1}{2} \gamma_{m} \quad (m = 0, 1, 2, 3), \quad  {\bf P}_{4} = - \frac{1}{2} \gamma_5.
\ee

\section{Relation between $\Theta$ and $r$-matrix for modified cYBE}
\label{sec:mCYBE}
In this section, we show that the NC structure identified in \cite{I-Sugra} for the ABF solution \cite{Arutyunov:2013ega, Arutyunov:2015qva} using the method outlined in section \ref{sec:open_string} agrees with the $r$-matrix. The analysis presented here extends this observation to YB deformations based on $r$-matrix solutions to the modified cYBE. We will restrict our attention to the bosonic generators of greatest interest to the geometry.

We begin by focusing on the bosonic part of the superalgebra $\frak{gl}(4|4)$ and in particular the conformal subalgebra, which entails restricting ourselves to the generators $E_{ij}$, where $i, j = 1, 2, 3, 4$. With this restriction, the commutation relations take the form \cite{Kawaguchi:2014qwa}:
\bea
\label{comm}
\left[ E_{ij}, E_{kl} \right] = \delta_{kj} E_{il} - \delta_{il} E_{kj}.
\eea

In this notation, dilatations $D$, translations $P_{\alpha \dot{\beta}}$ and special conformal transformations $K_{ \dot{\alpha} \beta}$, as well as a pair of $\frak{su}(2)$ subalgebras, $L_{\alpha \beta}$ and $\bar{L}_{\dot{\alpha} \dot{\beta}}$ take the form \cite{Kawaguchi:2014qwa}:
\bea
\label{conformal_gen}
D &=& \frac{1}{2} ( E_{\lambda \lambda} - E_{\dot{\lambda} \dot{\lambda}} ), \quad P_{\alpha \dot{\beta}} = E_{\alpha \dot{\beta}}, \quad K_{\dot{\alpha} \beta} = E_{\dot{\alpha} \beta}, \nn
L_{\alpha \beta} &=& E_{\alpha \beta} - \frac{1}{2} \delta_{\alpha \beta} E_{\lambda \lambda}, \quad \bar{L}_{\dot{\alpha} \dot{\beta}} = E_{\dot{\alpha} \dot{\beta}} - \frac{1}{2} \delta_{\dot{\alpha} \dot{\beta}} E_{\dot{\lambda} \dot{\lambda}},
\eea
where $\alpha, \beta, \lambda = 1, 2$ and $\dot{\alpha}, \dot{\beta}, \dot{\lambda} = 3, 4$.

To make contact with the geometry, we need to specify an AdS$_5$ metric and identify the Killing vectors associated to the above generators. Following ABF \cite{Arutyunov:2013ega, Arutyunov:2015qva}, we consider the metric,
\be
\dd s^2 = - ( 1+ \rho^2) \dd t^2 + \frac{\dd \rho^2}{(1+\rho^2)} + \rho^2 \left( \dd \zeta^2 + \cos^2 \zeta \dd \psi_1^2 + \sin^2 \zeta \dd \psi_2^2 \right).
\ee
It is a straightforward exercise to identify suitably complexified Killing vectors that satisfy the same commutation relations. Explicitly, the corresponding Killing vectors may be expressed as follows:
\bea
\label{Killing_vec}
D &=& - i \partial_t, \quad
L_{11}  = - \frac{i}{2} ( \partial_{\psi_1} + \partial_{\psi_2}), \quad
\bar{L}_{33} = - \frac{i}{2} ( \partial_{\psi_1} - \partial_{\psi_2}), \nn
L_{12} &=& e^{i ( \psi_1 + \psi_2)} \left( \tan \zeta \partial_{\psi_1} + i \partial_{\zeta} - \cot \zeta \partial_{\psi_2} \right), \nn
L_{21} &=& e^{- i ( \psi_1 + \psi_2)}  \left( \tan \zeta \partial_{\psi_1} - i \partial_{\zeta} - \cot \zeta \partial_{\psi_2} \right), \nn
\bar{L}_{34} &=& e^{i ( \psi_1 - \psi_2)} \left( \tan \zeta \partial_{\psi_1} + i \partial_{\zeta} + \cot \zeta \partial_{\psi_2} \right) , \nn
\bar{L}_{43} &=& e^{-i ( \psi_1 - \psi_2)} \left( \tan \zeta \partial_{\psi_1} - i \partial_{\zeta} + \cot \zeta \partial_{\psi_2} \right),
\eea
\bea
P_{14} &=& e^{i (t + \psi_1)} \frac{\sqrt{1+ \rho^2}}{\rho} \left( \rho \cos \zeta \partial_{\rho} + \frac{i \rho^2}{1 + \rho^2} \cos \zeta \partial_t  - \sin \zeta \partial_{\zeta} + i \sec \zeta \partial_{\psi_1} \right), \nn
P_{23} &=& e^{i (t - \psi_1)} \frac{\sqrt{1+ \rho^2}}{\rho} \left( \rho \cos \zeta \partial_{\rho} + \frac{i \rho^2}{1 + \rho^2} \cos \zeta \partial_t  - \sin \zeta \partial_{\zeta} - i \sec \zeta \partial_{\psi_1} \right), \nn
P_{13} &=& e^{i (t + \psi_2)} \frac{\sqrt{1+ \rho^2}}{\rho} \left( \rho \sin \zeta \partial_{\rho} + \frac{i \rho^2}{1 + \rho^2} \sin \zeta \partial_t  + \cos \zeta \partial_{\zeta} + i \csc \zeta \partial_{\psi_2} \right), \nn
P_{24} &=& e^{i (t -\psi_2)} \frac{\sqrt{1+ \rho^2}}{\rho} \left( \rho \sin \zeta \partial_{\rho} + \frac{i \rho^2}{1 + \rho^2} \sin \zeta \partial_t  + \cos \zeta \partial_{\zeta} - i \csc \zeta \partial_{\psi_2} \right), \nn
K_{41} &=& e^{-i (t + \psi_1)} \frac{\sqrt{1+ \rho^2}}{\rho} \left( \rho \cos \zeta \partial_{\rho} - \frac{i \rho^2}{1 + \rho^2} \cos \zeta \partial_t  - \sin \zeta \partial_{\zeta} - i \sec \zeta \partial_{\psi_1} \right), \nn
K_{42} &=& e^{-i (t -\psi_2)} \frac{\sqrt{1+ \rho^2}}{\rho} \left( \rho \sin \zeta \partial_{\rho} - \frac{i \rho^2}{1 + \rho^2} \sin \zeta \partial_t  + \cos \zeta \partial_{\zeta} + i \csc \zeta \partial_{\psi_2} \right), \nn
K_{31} &=& e^{-i (t + \psi_2)} \frac{\sqrt{1+ \rho^2}}{\rho} \left( \rho \sin \zeta \partial_{\rho} - \frac{i \rho^2}{1 + \rho^2} \sin \zeta \partial_t  + \cos \zeta \partial_{\zeta} - i \csc \zeta \partial_{\psi_2} \right), \nn
K_{32} &=& e^{-i (t - \psi_1)} \frac{\sqrt{1+ \rho^2}}{\rho} \left( \rho \cos \zeta \partial_{\rho} - \frac{i \rho^2}{1 + \rho^2} \cos \zeta \partial_t  - \sin \zeta \partial_{\zeta} + i \sec \zeta \partial_{\psi_1} \right),
\eea
where we have employed the same notation to make contact with the generators (\ref{conformal_gen}).

With the Killing vectors in hand, it is now an easy exercise to take the $r$-matrix from the literature, e. g. \cite{Kawaguchi:2014qwa}, and translate the generators into Killing vectors,
\bea
r &=& i c  \left( E_{12} \wedge E_{21} + E_{13} \wedge E_{31} + E_{14} \wedge E_{41} + E_{23} \wedge E_{32} + E_{24} \wedge E_{42} + E_{34} \wedge E_{43} \right), \nn
&=& 4 c \left( \rho \partial_t \wedge \partial_{\rho} + \tan \zeta \partial_{\zeta} \wedge \partial_{\psi_1} \right),
\eea
where $c$ is a complex constant. At this point, we have to simply recall the expression for $\Theta$ \cite{I-Sugra},
\be
\Theta^{t \rho} = \kappa \rho, \quad \Theta^{\zeta \psi_1} = \kappa \tan \zeta,
\ee
where $\kappa$ is an additional constant, to confirm that they are of the same form. It is clear that by normalizing Killing vectors correctly, one can recover (\ref{Theta-r}). Although we have focused exclusively on the deformation of the AdS$_5$ spacetime, the internal five-sphere is also deformed in the ABF solution. Either by analytic continuation, or direct calculation, it can be shown that the above result also extends to the five-sphere. We leave the exercise to the interested reader.

\section{Generalized type IIB supergravity equations}
\label{sec:gen_IIB}

In the case of Yang-Baxter deformations of $\5$ based
on the homogeneous CYBE~\cite{Kawaguchi:2014qwa}, the resulting target spacetime satisfies the equations of motion of type IIB supergravity if the classical $r$-matrix satisfies the unimodularity condition~\cite{Borsato:2016ose}. If not, the background is a solution of the generalized type IIB supergravity~\cite{Arutyunov:2015mqj,Wulff:2016tju} described below.

The generalised type IIB supergravity equations  are~\cite{Arutyunov:2015mqj,Wulff:2016tju}
\begin{gather}
R_{MN}-\frac{1}{4}H_{MKL}H_N{}^{KL}-T_{MN}+D_MX_N+D_NX_M=0 \qquad (M,N=0,1,\cdots,9)\,,
\label{general1} \\
\frac{1}{2}D^K H_{KMN}+\frac{1}{2}\mathcal{F}^K\mathcal{F}_{KMN}+\frac{1}{12}\mathcal{F}_{MNKLP}\mathcal{F}^{KLP} = X^K H_{KMN}+D_M X_N-D_N X_M\,, \label{general2}\\
R-\frac{1}{12}H^2+4D_MX^M-4X_MX^M=0\,,  \label{general3}\\
D^M\mathcal{F}_M-Z^M \mathcal{F}_M-\frac{1}{6}H^{MNK}\mathcal{F}_{MNK}=0\,,
\qquad I^M\mathcal{F}_M=0\,, \label{general4} \\
D^{K}\mathcal{F}_{KMN}-Z^K \mathcal{F}_{KMN}
- \frac{1}{6}H^{KPQ}\mathcal{F}_{KPQMN}
- (I\wedge \mathcal{F}_1)_{MN}=0\,, \label{general5} \\
D^{K}\mathcal{F}_{KMNPQ}-Z^K \mathcal{F}_{KMNPQ} +\frac{1}{36}\e_{MNPQRSTUVW}H^{RST}\mathcal{F}^{UVW}
- (I\wedge \mathcal{F}_3)_{MNPQ}=0\,.
\label{general6}
\end{gather}
The first equation (\ref{general1}) describes the dynamics of the metric
in the string frame $g_{MN}$, where $T_{MN}$ is defined as
\be\bn
T_{MN} &\equiv \frac{1}{2}\mathcal{F}_M\mathcal{F}_N
+\frac{1}{4}\mathcal{F}_{MKL}\mathcal{F}_N{}^{KL}
+\frac{1}{4\times 4!}\mathcal{F}_{MPQRS}\mathcal{F}_N{}^{PQRS} \\
&- \frac{1}{4}G_{MN} (\mathcal{F}_K\mathcal{F}^{K}
+\frac{1}{6}\mathcal{F}_{PQR}\mathcal{F}^{PQR})\,.
\en\ee
Here $\mathcal{F}_M\,,\mathcal{F}_{MNK}\,,\mathcal{F}_{MNKPQ}$ are
the rescaled RR field strengths
\be
\mathcal{F}_{n_1n_2\ldots}={\rm e}^{\Phi}F_{n_1n_2\ldots}\,,
\ee
where $\Phi$ is the dilaton whose motion is described by
(\ref{general3})\,.
The second equation (\ref{general2}) is for the field strength
$H_{MNK}$ of the NS-NS two-form.
Eqs. (\ref{general4}), (\ref{general5}) and (\ref{general6}) are respectively
for the R-R one-form, three-form and five-form field strengths. Note that $D_M$ is the covariant derivative with respect to the Levi-Civita connection of the closed string metric $g_{MN}$.

The Bianchi identities for the R-R field strengths are also generalized as
\begin{align}
&(\bd{\mathcal{F}_1} - Z\wedge \mathcal{F}_1)_{MN} - I^K\mathcal{F}_{MNK}=0, \\
&(\bd{\mathcal{F}_3} - Z\wedge \mathcal{F}_3 + H_3\wedge \mathcal{F}_1)_{MNPQ}
- I^K\mathcal{F}_{MNPQK} = 0, \\
&(\bd{\mathcal{F}_5} - Z\wedge \mathcal{F}_5 + H_3\wedge \mathcal{F}_3)_{MNPQRS}
+ \frac16\, \e_{MNPQRSTUVW}I^T\mathcal{F}^{UVW} = 0.
\end{align}

\medskip

Together with the standard type IIB fields,  equations (\ref{general1})-(\ref{general6})
involve the three new vector fields $X$, $I$ and $Z$. Let us consider them in detail.
In fact, only two of them are independent as the vector $X$ is expressed as
\be
X_M\equiv I_M+Z_M\,.
\ee
$I$ and $Z$ satisfy the following relations:
\be
D_M I_N+D_N I_M=0\,,\qquad D_M Z_N-D_N Z_M+I^K H_{KMN}=0\,, \qquad I^M Z_M=0\,.\label{IZ}
\ee
The first equation of (\ref{IZ}) is the Killing vector equation.
Assuming that $I_M$ is chosen such that the Lie derivative vanishes,
\be
(\mL_I B)_{MN}=I^K\partial_K B_{MN}+B_{KN}\partial_{M}I^K-B_{KM}\partial_N I^K=0,
\ee
the second equation of (\ref{IZ}) can be solved by
\be
Z_M=\partial_M\Phi-B_{MN}I^N\,.
\ee
Thus $Z$ can be regarded as a generalization of the dilaton gradient $\partial_M\Phi$\,.
In particular, when $I$ vanishes, $Z_M$ becomes $\partial_M\Phi$ and
the generalised equations (\ref{general1})-(\ref{general6}) are reduced
to the usual type IIB supergravity equations.

\subsection{$\Lambda$-invariance of generalised supergravity}
\label{sec:gen_IIB-inv}

Let us assume that the generalized supergravity equations of motion are coming from an action  $S_{g.s}$ \footnote{The explicit form of $S_{g.s.}$ is not known and it may not exist. However, here it is not important for our analysis, since it boils down to taking the divergence of an equation of motion of generalized supergravity and demanding consistency. The assumption of an action is merely a prop to aid presentation.}. We can ensure $\Lambda$-invariance of the theory (see section~\ref{sec:unimodular}) by the following simple argument. Assuming that the only spacetime field that transforms nontrivially under $\Lambda$-symmetry is the $B$ field with $\delta_\Lambda B = d \LL$, the variation of the action can be written as
\be
\d_\LL S_{g.s.} = \int d^{10} x\,\sqrt{g}\, \frac{\d \mathcal{L}_{g.s.}}{\d B_{MN}}\, \p_M \LL_N = - \int d^{10}x \, \sqrt{g}\, D_M Y^{MN} \LL_N,
\ee
where
\be
Y^{MN} =  \frac{\d \mathcal{L}_{g.s.}}{\d B_{MN}} = 0
\ee
is the generalised equation of motion for the $B$ field~\eqref{general2}. Checking that the divergence $D_M Y^{MN}$ of this expression vanishes we establish the $\LL$-invariance of the total generalised supergravity action. One can also view this as a simple consistency check for the complete set of modified supergravity field equations, as almost every equation is involved.

Using the equation of motion~\eqref{general2}:
\be\label{2}\bn
Y_{MN} &= \frac12 D^K H_{KMN} + \frac12 \mathcal{F}^K \mathcal{F}_{KMN} + \frac{1}{12} \mathcal{F}_{MNKLP} \mathcal{F}^{KLP} \\&- X^K H_{KMN} - D_M X_N + D_N X_M,
\en\ee
the divergence $D_M Y^{MN}$ splits into three parts:
\be
\frac12  D^M D^K H_{KMN} = -\frac14 R_{NKLM} H^{KLM} = 0.
\ee
By the symmetry of the Riemann tensor,
\be\label{4}\bn
D^M &\left( X^K H_{KMN} + D_M X_N - D_N X_M \right) \\&=- 2 I^K T_{NK} + 2 Z^K \left( \mathcal{F}^M \mathcal{F}_{MKN} + \frac16 \mathcal{F}_{KNMLP} \mathcal{F}^{MLP} \right),
\en\ee
where one has to use the identities
\be
D_M Z_N - D_N Z_M + I^K H_{KMN} = 0,\quad X_M = I_M + Z_M,
\ee
\eqref{2}, the modified Einstein equation
\be
R_{MN} - \frac14 H_{MKL} {H_N}^{KL} = T_{MN} -D_M X_N - D_N X_M,
\ee
and the fact that $I^M$ is a Killing vector (whence $D_M D_N I^K = {R^K}_{NML} I^L$). Finally, there is the divergence of the RR part of~\eqref{2}:
\be\label{7}
\frac12 D^M (\mathcal{F}^K \mathcal{F}_{KMN}) + \frac{1}{12} D^M (\mathcal{F}_{MNKLP} \mathcal{F}^{KLP}).
\ee
To evaluate this one has to use the modified equations of motion for the 1-form and the 3-form, as well as the Bianchi identities for the 3- and the 5-form. Adding up the contributions of~\eqref{4} and~\eqref{7} we find that they cancel, so that the total divergence is zero.

\section{Fermionic T-duality}
\label{sec:review_ftduality}
We recall the salient features of fermionic T-duality \cite{Berkovits:2008ic} (see \cite{OColgain:2012si} for a review), as it pertains to the supergravity description.  Firstly, for type II supergravity, one considers two Majorana-Weyl Killing spinors
\be
\label{spinor}
\eta = \left( \begin{array}{c} \e \\ \hat{\e} \end{array}\right),
\ee
whose chirality is a property of the supergravity. Fermionic T-duality is implemented through an auxiliary matrix $C_{ab}$, which satisfies
\be
\label{C}
\partial_{\mu} C_{ab} = 2 \, i \, \e^{T}_{a} \gamma_{\mu} \e_b,
\ee
where $\gamma_{\mu}$ are $16\times 16$ gamma matrices \footnote{Since we will be interested in type IIB geometries, $\e, \hat{\e}$ have the same chirality and once the chirality projection condition is imposed, the spinors have 16 independent components, which is reflected in the dimensionality of the gamma matrices.} and $\e_a$ are Killing spinors subject to the constraint:
\bea
\label{constraint}
\e^{T}_{a} \gamma_{\mu} \e_{b} + \hat{\e}_{a}^T \gamma_{\mu} \hat{\e}_b = 0.
\eea
Here $a = 1, \dots, n$, labels the number of fermionic T-dualities being performed. Once this condition is satisfied, the fermionic isometries commute \cite{Berkovits:2008ic}.

Our conventions mean that $ \gamma_0 \propto 1_{16}$, so that the Majorana (reality) condition on the spinors is at odds with (\ref{constraint}). As a result, we must complexify the spinors so that the constraint can be imposed. As a direct consequence, the supergravity solutions one generates through fermionic T-dualities are typically complex \cite{Bakhmatov:2009be}.

Once the matrix $C$ is determined, the transformation of the dilaton and the RR sector may be easily deduced from the following expressions \cite{Berkovits:2008ic}:
\bea
\label{fermion_dilaton} \tilde{\Phi} &=& \Phi + \frac{1}{2} \log \textrm{det} C,  \\
\label{fermion_RR} \frac{i}{16} e^{\tilde{\Phi} } \tilde{F}^{ij} &=& \frac{i}{16} e^{\Phi} F^{ij} + C_{ab}^{-1} \e_{a}^i \otimes \hat{\e}_{b}^j,
\eea
where for type IIB, we can define the RR sector bi-spinor
\be
F^{ij} = (\gamma^{\mu})^{ij} F_{\mu} + \frac{1}{3!} (\gamma^{\mu \nu \rho} )^{ij} F_{\mu \nu \rho} + \frac{1}{2} \frac{1}{5!} (\gamma^{\mu \nu \rho \sigma \lambda} )^{ij} F_{\mu \nu \rho \sigma \lambda}.
\ee
The metric does not change in the transformation. Although, the fermionic T-duality rules were initially derived using pure spinor formalism, they can be derived directly from supergravity via ansatz \cite{Godazgar:2010ph}. \footnote{See  \cite{Sfetsos:2010xa} for a treatment in terms of a canonical transformation}

\subsection{AdS$_5\times$S$^5$ self-T-duality}
From the analysis of \cite{Berkovits:2008ic}, we know that AdS$_5\times$S$^5$ is self-dual under a combination of four bosonic T-dualities along the translation symmetries of AdS$_5$ and eight compensating fermionic T-dualities. In contrast to  \cite{Berkovits:2008ic}, here our treatment will be slightly different, as we will impose the constraint (\ref{constraint}) and  directly solve (\ref{C}). Moreover, we will focus on fermionic T-dualities involving superconformal supersymmetries.

Solving the Killing spinor equations, \footnote{In this subsection we use the conventions of  \cite{Kelekci:2014ima}, which differ in an overall sign in the RR sector from  \cite{OColgain:2012ca}.} while dropping the Poincar\'e supersymmetries, we arrive at the solution:
\be
\e = \left( z^{\frac{1}{2}} + z^{-\frac{1}{2}} x^{\mu} \Gamma_{\mu}^{~4}  \right) e^{ \frac{\theta_1}{2} \Gamma^{6789} i \sigma^2} e^{\frac{\theta_2}{2} \Gamma^{56}} e^{\frac{\theta_3}{2} \Gamma^{67}}  e^{\frac{\theta_4}{2} \Gamma^{78}} e^{\frac{\theta_5}{2} \Gamma^{89}} \eta_S,
\ee
where $\eta_S$ is a constant spinor satisfying $\Gamma^{0123} (i \sigma^2) \eta_S = \eta_S$. Note, we have used the customary nested coordinates for S$^5$:
\be
\dd s^2 (S^5) = \dd \theta_1^2 + \sin^2 \theta_1 ( \dd \theta_2^2 + \sin^2 \theta_2 ( \dd \theta_3^2 + \sin^2 \theta_3 ( \dd \theta_4^2 + \sin^2 \theta_4 \dd \theta_5^2) ) ).
\ee
Imposing this projection condition, which relates the two Weyl spinors, we can extract sixteen-component Weyl spinors,
\be\bn
\e &= \left[ z^{\frac{1}{2}} + z^{-\frac{1}{2}} ( -x_0 \, \gamma^{4} + x_i \, \gamma_{i}^{~4} )\right] e^{ \frac{\theta_1}{2} \gamma^{45}} e^{\frac{\theta_2}{2} \gamma^{56}} e^{\frac{\theta_3}{2} \gamma^{67}}  e^{\frac{\theta_4}{2} \gamma^{78}} e^{\frac{\theta_5}{2} \gamma^{89}} \eta_S, \\
\hat{\e} &= - \left[ z^{\frac{1}{2}} + z^{-\frac{1}{2}} ( -x_0 \, \gamma^{4} + x_i \, \gamma_{i}^{~4} )\right] \gamma^{123} e^{ \frac{\theta_1}{2} \gamma^{45}} e^{\frac{\theta_2}{2} \gamma^{56}} e^{\frac{\theta_3}{2} \gamma^{67}}  e^{\frac{\theta_4}{2} \gamma^{78}} e^{\frac{\theta_5}{2} \gamma^{89}} \eta_S,
\en\ee
where repeated $i = 1, 2, 3,$ indices are summed.

Our task now is to identify the eight Killing spinors that satisfy the constraint (\ref{constraint}), on the further condition that all angular dependence drops out of the determinant, and as a result, the dilaton shift (\ref{fermion_dilaton}). If commuting isometries are chosen in an ad hoc way so that the angular dependence does not drop out of the dilaton shift, then additional internal T-dualities along (complexified) isometries in S$^5$, as explained initially in \cite{Berkovits:2008ic}, will be required. We expect that this procedure is suitably flexible, so that no matter how one chooses the eight commuting fermionic isometries, then one should always be able to find a compensating internal T-duality.

We recall that $\eta_S$, which appears in both $\e$ and $\hat{\e}$, is a sixteen-component spinor, so we should isolate eight Killing spinors corresponding to the fermionic isometries. Analysis of the constraint (\ref{constraint}) reveals that all angular dependence drops out and imposing it boils down to choosing the constant spinor $\eta_S$ so that
\be
\eta_S^{T} \gamma_{\mu} \eta_S = 0, \quad \mu = 0, 1, 2, 3.
\ee
Since $\gamma_{0}$ is proportional to the identity matrix, we can most easily identify eight Killing spinors by imposing the constraint $\gamma^{123}  \eta_S = i \eta_S$. In the process, our sixteen candidate Killing spinors are automatically reduced to eight \textit{complex} Killing spinors, with the constraint now satisfied by construction.

It is now straightforward to determine $C_{ab}$ by integrating (\ref{C}). Owing to the fact that the matrix is $8\times 8$, we omit the details. However, from the determinant, modulo an irrelevant constant, one can determine the shift in the dilaton,
\be
\Phi = 4 \log \left[ \frac{(-x_0^2 + x_1^2 + x_2^2 + x_3^2 +z^2)}{z} \right].
\ee
One next inverts $C_{ab}$ and determines the transformation of the RR sector from the bispinor (\ref{fermion_RR}). The output of the eight fermionic T-dualities takes the form:
\be\bn
\dd s^2 &= \frac{(- \dd x_0^2 + \dd x_1^2 + \dd x_2^2 + \dd x_3^2 + \dd z^2) }{z^2} + \dd s^2(S^5), \\
F_1 &=  \frac{4 i \, z^3}{(x_{\mu} x^{\mu} +z^2)^3} \dd \left[ \frac{z}{ (x_{\mu} x^{\mu} + z^2) } \right], \quad \Phi = 4 \log \left[ \frac{(x_{\mu} x^{\mu} +z^2)}{z} \right].
\en\ee
From here it is easy to see that the transformation (\ref{outer_auto_coord}), followed by bosonic T-duality along $x_0, x_1, x_2, x_3$ and an inversion $z \rightarrow z^{-1}$ restores the geometry to its original form.

\section{Cohomology classes for NC structure}
\label{appen:cohomology}

Having motivated the unimodularity condition (\ref{unimodular}) from a physical symmetry principle in the last section, here we briefly comment on a possible mathematical interpretation.
Despite the fact that $\Theta$ may not define a Poisson structure in the brane worldvolume,\footnote{To see the latter, let  $\S$ be a D-brane worldvolume with local coordinates $\xi^\m$, $\m=1, \cdots, d=p+1$ and define the $\Lambda$-invariant field ${\cal F}=(B+F)|_{\S}$ with field strength $H=\dd {\cal F}=\dd B$. The noncommutativity bivector \cite{Seiberg:1999vs, Szabo:2006wx} $\Theta=\frac{1}{2}\Theta^{\m\n}\partial_\m\wedge\partial _\n $ has components given by \eqref{Theta-def}
\be
\Theta^{\m\n}=- \left(\frac{1}{g+{\cal F}}{\cal F}\frac{1}{g-{\cal F}} \right)^{\m\n}\; .\nonumber
\ee
For odd $p$ and when  ${\cal F}$ is of maximal rank, and  when  $H=0$ then, ${\cal F}$ is a non-degenerate closed two-form and hence defines a sympletic structure on $\S$; $\S$ is hence a symplectic manifold \cite{Seiberg:1999vs,Szabo:2006wx}.  When ${\cal F}$ is degenerate, it does not define a symplectic structure. When $\dd {\cal F}\neq 0$ and the star product is defined through the Kontsevich formula (eq. 5.4 of \cite{Szabo:2006wx}), the associativity property of the star product implies that the $\Theta$ parameter defines a Poisson structure on the brane worldvolume \cite{Szabo:2006wx, Szabo:2005jj, Cornalba:2001sm}. On the other hand, in our case, we do not define the star-product through the Kontsevich formula, but by a Drinfeld twist, therefore, we do not need to impose any extra condition on the $\Theta$ parameter, and in particular, it does not necessarily define a Poisson structure on the brane worldvolume theory.} we can use the technology from the Poisson geometry to study some properties on the brane worldvolume. In particular, we can define a cohomology class for the NC structure, similar to the modular vector class we have in a generic Poisson manifold \cite{laurent2012, Kosmann2007}.

A Poisson manifold $(\S, \Theta)$ is a smooth manifold endowed with a skew-symmetric $2$-tensor $\Theta$, called a \emph{Poisson bivector} \cite{Kosmann2007}. With this structure, we can define the usual Poisson bracket $\{f, g\}:=\Theta^{\m\n}\partial_\m f\partial_\n g$ and the Hamiltonian vector field $H_f:=\{f, \cdot\}$.
Therefore, NC structure $\Theta$ may be used to define a Hamiltonian flow,
\be
H_f:=\Theta^{\m\n}\frac{\partial f}{\partial \xi^\m} \frac{\partial }{\partial \xi^\n} \; , \forall f \in {\cal C}(\S)\; .\label{hamil.vec}
\ee

There is a natural choice for the brane volume-form, that is
\be
\O=e^{-\phi}\sqrt{\det(g+{\cal F})}\dd^d \xi\; ,
\ee
as Szabo argues in \cite{Szabo:2006wx}. In this case, the Lie derivative of the volume-form is proportional to the volume-form itself, that is
\be
{\cal L}_X \Omega = X^\m \partial_\m\left(e^{-\phi}\sqrt{\det(g+{\cal F})} \right)\dd^d\xi\equiv h(\xi)\O\; ,
\ee
where $h(\xi)$ is a function. Furthermore, using  ${\cal L}_X:= \dd\circ \iota_X + \iota_X\circ \dd$, one can sow that
\begin{equation}
{\cal L}_X \Omega  =  \nabla\cdot X\; \Omega\; .
\end{equation}

Let us now define a vector field, which we call \emph{modular vector field} of $\Theta$ to follow the nomenclature of the Poisson manifold literature, by the condition
\be
\label{mod.vec02}
{\cal L}_{H_f}\Omega:= \chi(f)\Omega\;.
\ee
In other words, it is a vector field that  determines the Lie derivative along the Hamiltonian vector field (\ref{hamil.vec}). The LHS of (\ref{mod.vec02}) is
\begin{subequations}
\begin{equation}
{\cal L}_{H_f}\Omega = \dd f\wedge \dd \circ \iota_{\Theta}\O\; ,
\end{equation}
and the RHS is
\be
\chi(f)\Omega  = \chi^\m \d_\m f \ \Omega = \dd f \wedge \iota_{\chi} \Omega\; .
\ee
\end{subequations}
Therefore
\be
\iota_{\chi}\O=\dd\circ \iota_{\Theta}\ \O\; .\label{mod.vec2}
\ee

The divergence $\nabla\cdot \Theta$ reads
$$
\star\nabla\cdot \Theta = \dd \star \Theta =  \dd (\star \Theta)= \dd \left(\iota_\Theta \O \right)
$$
and using (\ref{mod.vec2}), one can show that $\nabla\cdot\Theta = \chi$.
On the other hand, we have already shown  $ \nabla\cdot\Theta = I|_\S$.
So, the vector $I|_\S$ defines the modular vector field for the NC structure $\Theta$.

There is one interesting consequence of this definition. Let us suppose that there exists a second volume form $\widehat{\O}$ such that $\O=g(\xi)\widehat{\O}$. Therefore
\begin{align}
\chi(f) \O & = {\cal L}_{H_f}\O = {\cal L}_{H_f}\left( g(\xi)\widehat{\O} \right)\nn
& = H_f (g)\widehat{\O}+ g {\cal L}_{H_f} (\widehat{\O})
=-H_g(f) \widehat{\O} +g \hat{\chi} \widehat{\O}\; .
\end{align}
Then,
\be
\chi=\hat{\chi}-\frac{1}{g}H_g\;.
\ee
One can use this equivalence relation to define a cohomology class $[\chi]\in H^1(\S)$ \cite{Kosmann2007, Caseiro2011, Ciccoli2009, laurent2012}. The important point is that, when the cohomology class of modular forms is trivial $[\chi]=0$, then we can take a representative element of $[\chi]$ such that $\chi=H_F$ (equivalent to a closed form in the de Rham cohomology) and in this case, we have
\be
\begin{split}
{\cal L}_{H_f} (e^F\Omega) & = H_f(e^F)\O + e^F {\cal L}_{H_f}\O\\
& = - e^F H_F(f)\O + e^F H_F(f)\O = 0\; .
\end{split}
\ee
Therefore, in principle, $I^\m$  need not be a constant, as we already know from the generalised supergravity solutions, but we can demand that it belongs to the trivial cohomology class of the NC structure $\Theta$, that is,
	\be
	I^\m=\Theta^{\m\n}\partial_\n F\; .
	\ee
This requirement may be understood noting the existence of a nontrivial modular class which implies that we do not have a volume form $\O$ invariant under all Hamiltonian flows. A similar feature may be encountered in classical mechanics, but in this case the Liouville theorem states that the volume form in a sympletic manifold is preserved by Hamiltonian flows. In other words, Hamiltonian flows preserve volumes in the phase space \cite{arnold1989, frankel2003}.

%
%
%\setstretch{1.0}
%\bibliographystyle{utphys}
%\bibliography{bibfile}

\providecommand{\href}[2]{#2}\begingroup\raggedright

\end{document}